\documentclass[12pt]{emulateapj}

\submitted{To appear in ApJ}
\shorttitle{Weak Lensing by Large-Scale Structure with {\it FIRST}}
\shortauthors{Chang, Refregier \& Helfand}

\begin{document}
\title{Weak Lensing by Large-Scale Structure with the FIRST Radio 
Survey}

\author{Tzu-Ching Chang$^{1,2}$, Alexandre Refregier$^{3,4}$, David
J. Helfand$^1$}
\affil{1 Department of Astronomy and Columbia Astrophysics Laboratory,
Columbia University, 550 W. 120th Street, New York, NY 10027}
\affil{2 Department of Astronomy, University of California at Berkeley, 601 
Campbell Hall, Berkeley, CA 94720; tchang@astro.berkeley.edu} 
\affil{3 Service d'Astrophysique, CEA/Saclay, 91191 Gif sur Yvette, France}
\affil{4 Institute of Astronomy, Madingley Road, Cambridge CB3 OHA, UK}

\begin{abstract}
We present the first measurement of weak lensing by large-scale
structure on scales of $1-4$ degrees based on radio observations.  We
utilize the FIRST Radio Survey, a quarter-sky, 20~cm survey produced
with the NRAO Very Large Array (VLA). The large angular scales
afforded by the FIRST survey provide a measurement in the linear
regime of the matter power spectrum, thus avoiding the necessity of
applying uncertain non-linear corrections.  Moreover, since the VLA
interferometer has a well-known and deterministic beam, our
measurement does not suffer from the irreproducible effects of
atmospheric seeing which limit ground-based optical surveys. We use
the shapelet method described in an earlier paper to estimate the
shear from the shape of radio sources derived directly from the
interferometric measurements in the Fourier($uv$) plane.  With
realistic simulations we verify that the method yields unbiased shear
estimators.  We study and quantify the systematic effects which can
produce spurious shears, analytically and with simulations, and
carefully correct for them.  We measure the shear correlation
functions on angular scales of $0.5^{\circ}-40^{\circ}$, and compute
the corresponding aperture mass statistics. On $1^{\circ}-4^{\circ}$
scales, we find that the B-modes are consistent with zero, and detect
a lensing E-mode signal significant at the 3.0$\sigma$ level.  After
removing nearby radio sources with an optical counterpart, the
$E$-mode signal increases by 10-20\%, as expected for a lensing signal
derived from more distant sources. We use the E-mode measurement on
these scales to constrain the mass power spectrum normalization
$\sigma_8$ and the median redshift $z_m$ of the unidentified radio
sources. We find $\sigma_8 (z_m/2)^{0.6} \simeq 1.0\pm0.2$ where the
$1\sigma$ error bars include statistical errors, cosmic variance, and
systematics. This is consistent with earlier determinations of
$\sigma_{8}$ from cosmic shear, the cosmic microwave background (CMB)
and cluster abundance, and with our current knowledge of the redshift
distribution of radio sources. Taking the prior $\sigma_{8}=0.9\pm0.1$
(68\%CL) from the WMAP experiment, this corresponds to
$z_{m}=2.2\pm0.9$ (68\%CL) for radio sources without optical
counterparts, consistent with existing models for the radio source
luminosity function. Our results offer promising prospects for
precision measurements of cosmic shear with future radio
interferometers such as LOFAR and the SKA.
\end{abstract}

\section{Introduction}

Weak gravitational lensing by large-scale structure, or `cosmic
shear', has emerged as a powerful tool for measuring the mass
distribution of the Universe (see van Waerbeke \& Mellier 2003;
Refregier 2003 for recent reviews). This effect is based on the
distortion induced in the images of background galaxies by the
gravitational tidal field of the intervening large-scale structure.
Since light rays from nearby background galaxies travel through
similar mass inhomogeneities along the way, the resulting image
distortions are correlated.  By measuring the coherent lensing
distortions of background galaxies, one can thus probe the mass
distribution projected along the line of sight. The underlying physics
is well-understood, and the observational signatures can be directly
compared to theoretical predictions. The distortion depends on the
mass fluctuations, as well as on the mean mass density of the
Universe.  Since it is a projected effect, the amplitude is also
sensitive to the geometry of the Universe and on the background source
distribution in redshift space.

The lensing distortion can be quantified by a $2 \times 2$ tensor
field, parameterized by the convergence and shear parameters, with the
latter being the direct observable.  Since galaxies have an intrinsic
shape distribution, the shear signal must be measured statistically,
assuming galaxies are on average randomly oriented. The lensing effect
of interest here is not associated with any specific mass
concentrations, but is measured by sampling random regions of the
sky. The typical amplitude of the shear in this regime is only on the
order of a few percent on arcminute scales, and is therefore
challenging to measure. Large areas of the sky must be surveyed to
reduce statistical errors, and systematic effects that spuriously
distort the shape of the observed sources need to be carefully
corrected for.

The theoretical basis for cosmic shear was pioneered by Gunn (1967),
and was further elaborated in a modern cosmology language by, e.g.,
Blandford et al. (1991), Miralda-Escude (1991), Kaiser (1992, 1998),
Jain and Seljak (1997), and Schneider (1998). Because of its
challenging observational nature, cosmic shear was firmly detected for
the first time only recently (Bacon et al. 2000; Kaiser et al. 2000;
van Waerbeke, et al. 2000; Wittman et al. 2000).  Since then, the
cosmological origin and implications of the shear signal have been
rapidly established by many more groups (most recently Bacon et al
2003; Brown et al. 2003; Hamana et al. 2003; Hoekstra et al. 2002;
Jarvis et al. 2003; Refregier et al. 2002; van Waerbeke et al. 2002).
Collectively, the shear two-point statistics have been measured from
$\theta \sim 1'$ to $100'$ angular scales, spanning the non-linear
($\theta \lesssim 10'$) and quasi-linear ($\theta \gtrsim 10'$) regime of
the matter power spectrum, and are consistent with predictions from
the popular $\Lambda$CDM model.  Constraints on the matter density
$\Omega_{m}$ and the power spectrum normalization $\sigma_{8}$ from
different cosmic shear measurements are broadly in agreement with one
another, and are roughly consistent with measurements from the cluster
abundance method (e.g., Pierpaoli et al. 2001; Seljak 2002).

Until now, all current cosmic shear measurements have been performed
in the optical and near-infrared bands, with similar observational and
analysis techniques.  Here, we report a cosmic shear measurement
using the {\it FIRST} Radio Survey (Becker et al. 1995; White et
al. 1997).  {\it FIRST} is a quarter-sky radio survey at 1.4
GHz conducted using the Very Large Array (VLA) in its B-configuration, and
therefore provides a unique measurement of the mass power spectrum on
large angular scales (Kamionkowski et al. 1998; Refregier et al. 1998;
Chang \& Refregier 2002).  Radio surveys like {\it FIRST} offer
several advantages for cosmic shear. Firstly, unlike optical galaxies,
bright radio sources are typically at high redshifts, thereby
increasing the path-length to each through the Universe, and thus the
strength of the lensing signal. Secondly, the VLA is a
radio interferometer and thus has a well-known and deterministic beam,
allowing for modeling of crucial systematic effects to high
accuracy. In contrast, ground-based optical surveys are limited by the
irreproducible effects of atmospheric seeing. Finally, {\it FIRST} is
a sparsely-sampled but wide-angle survey, well-suited for measurements
in the linear part of the mass power spectrum. This avoids the
theoretical uncertainties arising from the non-linear corrections of
the mass power spectrum required to interpret the optical cosmic shear
surveys that are sensitive to smaller scales.

To deal with the important issue of accurate shear measurements and
systematics corrections, we use the shapelet method described in Chang
\& Refregier (2002, hereafter CR02; see also Refregier 2003b; Refregier
\& Bacon 2003) to measure the shape and shear information from
interferometric data directly in Fourier space.  The method is linear
and yields unbiased shear estimators.  Systematic effects such as
instrumental distortions can be closely modeled.

Our paper is organized as follows. In \S\ref{formalism} we describe
the theoretical background and notations for cosmic shear. In
\S\ref{data}, we summarize the main features of the FIRST radio survey
and the properties of its radio sources. In \S\ref{method}, we
describe briefly our shape measurement method, discussing the impact
of systematic effects in \S\ref{systematics}. The systematics
corrections are described in \S\ref{sys_corrections}.  In \S\ref{results},
we present our results for the measurement of cosmic shear, and
discuss their cosmological implications. Our conclusions are
summarized in \S\ref{conclusion}.

\section{Theory}
\label{formalism}

We first briefly describe the theoretical basis of cosmic shear and
the statistics we will use to present our results (see, e.g., Bartelmann
\& Schneider 2000 for a review).

\subsection{Weak Lensing}
In a weakly inhomogeneous FRW Universe, light rays from distant
sources are deflected by the gravitational tidal field of intervening
structures.  In the weak-lensing limit, the resulting distortion and
magnification can be quantified by a $2 \times 2$ tensor field,
$\psi_{lm}$, which, following from the geodesic equation, can be
related to the gravitational potential $\Phi$:
\begin{equation}
\label{eqn:potential}
\psi_{lm} = {2 \over c^2} \int d\chi g(\chi) \partial_{l} \partial_{m} \Phi
\end{equation}
where the radial window function $g(\chi)$ is defined as
\begin{equation}
\nonumber
g(\chi) = r(\chi) \int_{\chi}^{\chi_{h}} d\chi' n(\chi') {r(\chi'-\chi) \over r(\chi')}. 
\end{equation}
Here, $n(\chi')$ is the normalized source redshift distribution,
$r(\chi)$ is the comoving angular-diameter distance, $\chi$ is the
radial comoving coordinate, and $\chi_h$ corresponds to the
horizon. The comoving derivatives $\partial_{i}$ are perpendicular to
the line-of-sight. The symmetric distortion tensor $\psi_{lm}$ can be
parameterized by the convergence, $\kappa$, and the shear $\gamma_{1}$
and $\gamma_{2}$. These are defined as
\begin{eqnarray}
\label{eqn:kappa}
\kappa&=&{1 \over 2} (\psi_{11}+\psi_{22}) \\ 
\gamma_{1}&=&{1 \over 2} (\psi_{11}-\psi_{22}),
~~~\gamma_{2}=\psi_{12}=\psi_{21}       
\end{eqnarray}
Adding to Equation (\ref{eqn:kappa}) a $\psi_{33}$ term which cancels
out upon $\chi$ integration, and using Poisson's equation in the form
of $\nabla^{2}\Phi={3H_{0}^{2}\Omega_m \over 2a} \delta$, one relates
the convergence field $\kappa$ to the density fluctuation $\delta$,
weighted by the window function $g(\chi)$. Applying Limber's equation
in Fourier-space, the power spectrum of convergence, $P_{\kappa}$, can
then be expressed in terms of the 3-D mass power spectrum,
$P_{\delta}$ (e.g., Kaiser 1992) as
\begin{equation} 
P_{\kappa}(l)={9H_{0}^{4}\Omega_{m}^{2} \over 4c^{4}}
\int_{0}^{\chi_h} d\chi \left[{g(\chi) \over a(\chi)r(\chi)}
\right]^{2} P_{\delta} \left({l \over r(\chi)};\chi \right)
\end{equation}
where $\delta$ is the mass-density fluctuation, $a(\chi)$ is the
cosmological scale factor, and $H_0$ and $\Omega_m$ are the
present-day Hubble constant and  matter density parameter,
respectively. Note that the convergence and shear fields are related
by (e.g., Kaiser \& Squires 1993)
\begin{equation}
\kappa=\partial^{-2} \partial_i \partial_j \gamma_{ij},
\end{equation}
where $\partial^{-2}$ is the inverse 2D Laplacian operator
\begin{equation}
\partial^{-2} \equiv {1 \over 2\pi} \int d^2\hat{r} \ln|\hat{r}-\hat{r'}|.
\end{equation}
In the flat sky approximation, the convergence power spectrum is equal
to the shear power spectrum $P_\gamma = P_\kappa$.  The shear power
spectrum is simply related to other 2-point shear statistics
which are more convenient in practice and which we describe below.

\subsection{Shear Correlation Functions}
For each pair of galaxies, we define the tangential and
$45^{\circ}$-rotated shear components, $\gamma_t$ and $\gamma_r$, with
respect to the great circle connecting the two galaxies: as
\begin{eqnarray}
\nonumber
\label{eqn:sheart}
\gamma_t &=& \gamma_1 \cos(2\theta) + \gamma_2 \sin(2\theta) \\
\gamma_r &=& \gamma_2 \cos(2\theta) - \gamma_1 \sin(2\theta),
\end{eqnarray}
where $\theta$ is the position angle between the x-axis and the great
circle.  In analogy to the CMB polarization correlations, one can then
construct coordinate-independent correlations from the rotated shear
components (Kamionkowski et al. 1998).  There are three independent
pair-wise shear correlation functions, $C_1(\theta) =
\langle \gamma_t(\theta_0) \gamma_t(\theta_0+\theta) \rangle$, $C_2(\theta) =
\langle \gamma_r(\theta_0) \gamma_r(\theta_0+\theta) \rangle$, and $C_3(\theta) =
\langle \gamma_t(\theta_0) \gamma_r(\theta_0+\theta) \rangle$.  The first two
correlation functions are related to the shear power spectrum
(e.g., Miralda-Escude 1991)
\begin{eqnarray}
\nonumber
C_1(\theta)&=&\int {ldl \over 4\pi} P_\gamma(l)
[J_0(l\theta)+J_4(l\theta)] \\
C_2(\theta)&=&\int {ldl \over 4\pi} P_\gamma(l)
[J_0(l\theta)-J_4(l\theta)],
\end{eqnarray}
while the parity invariance of weak lensing ensures that $C_3(\theta)$
vanishes. The measurement of a non-zero $C_3(\theta)$ is, therefore,
an indication of a non-lensing contribution, such as that from residual
systematics.

\subsection{$M_{ap}$ Statistics}
The aperture mass is defined as (Kaiser 1995; Schneider et al. 1998)
\begin{equation}
M_{ap}(\theta)=\int_{|{\mathbf \phi}| < \theta} d^2{\mathbf \phi} U(|{\mathbf \phi}|)\kappa(\phi),
\end{equation}
where $U(\phi)$ is a compensated filter defined so that
$\int_0^{\theta} d\phi ~\phi U(\phi) =0$, and $\kappa(\phi)$ is the
convergence field. The $M_{ap}$ statistic thus measures the
spatially-filtered projected density field, and can be conveniently
related to the shear as
\begin{equation}
M_{ap}(\theta)=\int_{|{\mathbf \phi}| < \theta} d^2 {\mathbf \phi} Q(|{\mathbf \phi}|) \gamma_t(\phi),
\end{equation}
where $Q(\phi) = {2 \over \phi^{2}} \int_0^{\phi} d\phi' \phi'
U(\phi') - U(\phi)$, and $\gamma_t$ is the rotated tangential shear,
as in Eq.(\ref{eqn:sheart}), with respect to the aperture center.  The
aperture mass variance is related to the shear power spectrum by
(Schneider et al. 1998):
\begin{equation}
\langle M_{ap}^{2}(\theta) \rangle={288 \over \pi \theta^4} \int {dl \over l^3}
P_\gamma(l) [J_4(l\theta)]^2.
\end{equation} 

The $M_{ap}$ statistic has the advantage that the mass density can
be directly obtained from the observables -- the shear -- without the need for
mass reconstruction.  The weight function $Q$ is relatively narrow in
Fourier space, so that the $M_{ap}(\theta)$ measurements do not
strongly correlate.  As a result, $M_{ap}(\theta)$ and
$M_{ap}(2\theta)$ are almost independent of each other. Note that
$M_{ap}(\theta)$ probes the scale of $\sim \theta / 4$ (Schneider
1998).

Weak gravitational lensing arises from scalar perturbations to the
space-time metric, and therefore the shear field is expected to
possess no handedness.  One can decompose the shear tensor field into
gradient (E-mode) and curl (B-mode) components (Stebbins 1996). The
gradient part contains the weak lensing signal, while the curl part is
expected to be zero.  Gravitational waves produce non-zero B-mode
signals, but their effect is expected to be very small.  Thus, the B-mode
provides a useful check for any non-lensing contaminations, such as
residual systematic effects or intrinsic shape correlations (e.g., Heavens
2001). A rotation of 45 degrees in the aperture brings a curl-free
field to a curl field, and also transforms a tangential shear to a
radial shear component. Thus, the B-mode of the aperture mass,
$M_{\bot}$, is simply
\begin{equation}
M_{\bot}(\theta)=\int_{|{\mathbf \phi}| < \theta} d^2{\mathbf \phi} Q(|{\mathbf \phi}|) \gamma_r(\phi),
\end{equation}
where $\gamma_r(\phi)$ is the radial shear with respect to the
aperture center.  The aperture mass statistic therefore provides a
convenient way for E- and B-mode decomposition, and can be computed
directly from the observed shear.

The aperture mass variance can also be expressed in terms of the shear
correlation functions.  Defining
\begin{equation}
\label{eqn:c+}
C_+(\theta)=C_1(\theta)+C_2(\theta); ~~C_{-}(\theta)=C_1(\theta)-C_2(\theta),
\end{equation}
the aperture masses are given by
\begin{eqnarray}
\label{eqn:map}
\nonumber 
\langle M_{ap}^2(\theta) \rangle &=& {1 \over 2} \int_0^{2\theta} {\phi d\phi \over
\theta^2} [C_+(\phi) T_+({\phi \over \theta}) + C_{-}(\phi)
T_{-}({\phi \over \theta})] \\ \nonumber
\langle M_{\bot}^2(\theta) \rangle &=& {1 \over 2} \int_0^{2\theta} {\phi d\phi \over
\theta^2} [C_+(\phi) T_+({\phi \over \theta}) - C_{-}(\phi)
T_{-}({\phi \over \theta})],\\
\end{eqnarray}
where the functions $T_+$ and $T_{-}$ are defined in Schneider, van
Waerbeke \& Mellier (2002).

\section{Data}
\label{data}

\subsection{FIRST Survey}
\label{survey}
The FIRST radio survey (Becker et al. 1995; White et al. 1997) was
conducted at the NRAO Very Large Array (VLA) at 1.4 GHz in the B
configuration. The survey is the radio equivalent of the Sloan Digital
Sky Survey covering $\sim 10,000$ deg$^{2}$ of the Northern Galactic Cap.  It
consists of 3-minute snapshots covering a hexagonal grid using 14
3-MHz frequency channels. Its $5\sigma$ flux density limit is 1 mJy, with a
restoring beam FWHM of $5''.4$.  The survey contains $\sim 9\times
10^5$ sources, roughly 40\% of which are resolved.  The basic source
information is described in the on-line FIRST catalog
(http://sundog.stsci.edu/).

\subsection{Redshift Distribution}
\label{zdist}
For our weak lensing measurement, the source redshift distribution
determines the weight function along the line of sight, and thus
conditions the strength of the lensing signal.  The signal depends
most strongly on the source median redshift $z_m$ and somewhat on the
redshift distribution.  As discussed below, the redshift distribution
for radio sources is rather uncertain, and we have therefore
considered $z_m$ as a free parameter throughout the paper.

For our purposes, we consider the redshift distribution for radio
sources estimated by Dunlop \& Peacock (1990; hereafter DP). These
authors estimated the radio luminosity functions (RLF) for the steep-
and flat-spectrum sources at 2.7 GHz, using faint sources with optical
counterparts in the UGC catalog and the Parkes selected regions
database with source flux densities $> 100$ mJy. They presented seven redshift
models, with model 7 considered with two cases:  mean- and
high-redshift source distributions. The former have assigned redshifts
that are the mean values for galaxies of a given K-band luminosity,
and the latter have redshifts larger than the average, to account for
a possible bias. 

\begin{figure}[h]
\centering
\includegraphics[scale=0.42]{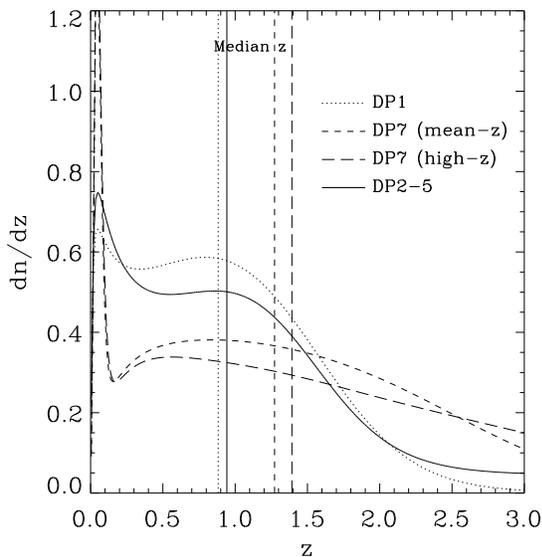}
\figcaption{Normalized differential redshift distribution of the FIRST
radio sources using the models by Dunlop and Peacock (1990, DP). 
The DP model 1, the average of models 2-5, and models 7 with the mean and high-redshift
estimations are shown. The vertical
lines indicate the median redshifts of each model.
\label{fig:dndz}
}
\end{figure}

The normalized differential redshift distribution of the DP models is
shown in Fig. \ref{fig:dndz}.  The RLF's were integrated over
luminosities that produce flux densities greater than 1 mJy. The DP
models were shifted to 1.4 GHz for computation, assuming a spectral
index of $\alpha = -0.85$ for steep-spectrum sources and $\alpha=0$
for flat-spectrum sources, where $L \propto \nu^{\alpha}$.  The DP
model 1 is the fundamental model, and models 2 to 5 are variations
relative to it. For clarity, only the averaged values of models 2 to 5
are shown.  The high redshift end was truncated at $z=4.5$, as models
4 and 5 have unphysical spikes at $z \sim 6$.  Model 7 includes
effects of source evolution in luminosity and density, and is
considered the most probable approximation (Magliocchetti et
al. 1999). The spike near $z=0$ is due to starburst galaxies; however,
optical follow-up observations suggest the spike is probably
unphysical (Magliocchetti et al. 2000).  The vertical bars show the
median redshifts for each model (truncated at $z=4.5$) which range
from $z_m \sim 0.9$ to $1.4$.

\section{Shear Measurement}
\label{method}

\subsection{Shapelet Method}
\label{shapelets}
We summarize the relevant components of the shapelet method
described in CR02.  In this
approach, the surface brightness $f({\mathbf x})$ of an object is
decomposed as 
\begin{equation}
\label{eq:cartesian}
f({\mathbf x}) = \sum_{{\mathbf n}} f_{{\mathbf n}} B_{{\mathbf n}}
({\mathbf x};\beta),
\end{equation}
where the Fourier transform of the basis functions are given by
\begin{equation}
B_{{\mathbf n}}({\mathbf x};\beta) \equiv 
\frac{H_{n_{1}}(\beta^{-1} x_{1}) ~H_{n_{2}}(\beta^{-1} x_{2})
   ~e^{-\frac{|x|^{2}}{2 \beta^{2}}}}
  {\left[ 2^{(n_{1}+n_{2})} \pi ~\beta^{2}
  ~n_{1}! ~n_{2}! \right]^{\frac{1}{2}}}
\end{equation}
representing the two-dimensional orthonormal Gauss-Hermite basis functions of
characteristic scale $\beta$, where $H_{m}(\xi)$ is the Hermite polynomial
of order m, ${\mathbf x}=(x_{1},x_{2})$ and ${\mathbf
n}=(n_{1},n_{2})$. In practice, the series only include a finite
number of terms with order $n_{1}+n_{2} \leq N_{\rm max}$.

The decomposition and its applications have also been studied by
Refregier (2003), Refregier \& Bacon (2003) and independently by
Bernstein \& Jarvis (2002). The bases are complete and yield fast
convergence in the expansion if $\beta$ and $x=0$ are close to the
size and location of the object, respectively.  The basis functions
can be thought of as perturbations around a two-dimensional Gaussian,
so that the coefficients are the Gaussian-weighted moments of the
source. The basis functions are also the eigenfunctions of the Quantum
Harmonic Oscillator. Under a linear coordinate transformation, such
as a weak shear, the basis functions have analytic responses, making
an unbiased shear estimator easy to construct from the coefficients.
Furthermore, the basis functions are their own Fourier transforms up
to a rescaling factor (see Eq.~\ref{eqn:duality} below), and are thus
localized functions in both real and Fourier spaces. They are thus
convenient for analyzing data from interferometers, which measure the
Fourier-transform of the sky surface brightness.

Making use of the basis functions, we model the source intensity
directly in the Fourier (or $uv$) space, where the FIRST
interferometric data (visibilities) are collected.  The Fourier
transform $\tilde{f}_{s}({\mathbf k})$ of object intensity
$f_{s}(l,m)$ of each source $s$ is a sum of the Fourier shapelets
basis functions
\begin{equation}
\tilde{f}_{s}({\mathbf k}) = \sum_{\mathbf n} f_{{\mathbf n}s} \tilde{B}_{\mathbf
n}({\mathbf k}-{\mathbf k}_{s};\beta_{s}^{-1}),
\end{equation}
where
\begin{equation}
\label{eqn:duality}
\widetilde{B}_{{\mathbf n}}({\mathbf k};\beta^{-1}) = i^{(n_{1}+n_{2})}
B_{{\mathbf n}}({\mathbf k};\beta).
\end{equation}
A $\chi^2$ fit is used to simultaneously
model all sources in the field:
\begin{equation}
\label{eq:chi2}
{\chi}^{2}=({\mathbf V} - {\mathbf M ~f})^{T}~{\mathbf
C}^{-1}~({\mathbf V} - {\mathbf M~f}),
\end{equation}
where ${\mathbf V}=\{ \overline{V}_{c} \}$ is the binned visibility
data vector, ${\mathbf M}=\{ \overline{B}_{c}^{{\mathbf n}s} \}$ is
the theory matrix whose components are the basis functions of
different order ${\mathbf n}$ evaluated at the respective source
${\mathbf k}_{s}$, and ${\mathbf f}=\{ f^{{\mathbf n}s} \}$ is the
coefficient vector for all sources.  We assume the data error matrix
${\mathbf C}$ is diagonal and constant, ${\mathbf C} = \sigma^2
{\mathbf I}$, where $\sigma$ is the noise level and ${\mathbf I}$ the
identity matrix; this is a reasonable approximation for the VLA
observations since the noise of antennas is independent.  To find the
best-fit solution, we solve a set of simultaneous least-squares
equations. The covariance matrix of the best-fit coefficients
${\mathbf f}$ is then
\begin{equation}
W({\mathbf f})=\sigma^2 ({\mathbf M}^{T}{\mathbf M})^{-1},
\end{equation}
which provides an estimate of the errors on the coefficients and can
be used to compute the errors of derived quantities, such as source flux
density and size, and the shear estimator. In \S\ref{params} below, we explain
how the shapelet centroid, scale $\beta$, and maximum order $N_{\rm
max}$ are chosen in practice.

An unbiased shear estimator for each source and for each
shapelet order $n$ is then given by
(Refregier \& Bacon 2003)
\begin{equation}
\label{eqn:shear}
\hat{\gamma}_{n}=\frac{4}{[n(n+2)]^{\frac{1}{2}}}
\frac{\hat{f}_{n,2}}{\langle \hat{f}_{n-2,0}-\hat{f}_{n+2,0} \rangle}
\end{equation}
where $\hat{\gamma}=\gamma_1+i\gamma_2$ is the complex shear, and the
brackets denote an average over an unlensed object ensemble.  The
coefficients $\hat{f}_{n,m}$ are the complex polar shapelet
coefficients which are related to the Cartesian coefficients of
equation~(\ref{eq:cartesian}) by a simple linear transformation (see
Refregier 2003b). In this paper, we will only consider the $n=2$ shear
estimator $\hat{\gamma}_{2}$, which captures most of the shape
information of faint radio sources. The error on the shear estimator
can be straight-forwardly propagated from the covariance matrix.

\subsection{Choice of Shapelet Parameters}
\label{params}
One issue which needs to be resolved is the choice of the shapelet
parameters -- $\beta$, $N_{\rm max}$ and centroid -- for each source.  The
shapelets centroid was simply chosen to match the source centroid
position in R.A. and Dec. listed in the FIRST catalog, converted into
$l,m,n$ coordinates. 

To generate this conversion, we note that the R.A. and Dec. coordinate
system, also called the $x,y,z$ coordinate system, is fixed to the 
Earth. It is defined so that the $x,y$ axes define the plane of the
equator and the $z$ axis points along the rotation axis of the
Earth. The $l,m,n$ system is identical to the $u,v,w$ system, and is
defined, for a given pointing, so that the $w$ axis points towards the
phase tracking center and the $u$ axis is in the $x,y$ plane. Let $h$
and $\delta$ be the hour angle and declination of a source. Its
coordinate in the $x,y,z$ system is then $x=\cos \delta \cos h, y=-
\cos \delta \sin h , z= \sin \delta$. The corresponding coordinates in
the $u,v,w$ coordinate system are then given by
\begin{eqnarray}
\left( \begin{array}{c} l \\ m \\ n \end{array} \right) & = &
\left( \begin{array}{ccc} \sin h_{0} & \cos h_{0} &
0 \\ - \sin \delta_{0} \cos h_{0} & \sin \delta_{0} \sin h_{0} & \cos
\delta_{0} \\ \cos \delta_{0} \cos h_{0} & - \cos \delta_{0} \sin
h_{0} & \sin \delta_{0} \\ \end{array} \right) \nonumber \\
 & & \times \left( \begin{array}{c}
 \cos \delta
\cos h \\ - \cos \delta \sin h \\ \sin \delta \end{array} \right),
\end{eqnarray}
where $h_{0}$ and $\delta_{0}$ are the hour angle and declination of
the phase tracking center for this pointing. If the source is at the
phase tracking center ($h=h_{0}, \delta=\delta_{0}$), the position on
the $u,v,w$ system is ${\mathbf l}=(0, 0, 1)$, as expected since
$l,m,n$ are directional cosines such that $l^2+m^2+n^2=1$. For small
displacements $h=h_{0}+\Delta h, \delta=\delta_{0} + \Delta \delta$
away from the phase tracking center, the $l,m$ position takes the
familiar form ${\mathbf l} \simeq {\mathbf l_{1}} = ( - \Delta h \cos
h_{0}, \Delta \delta, 1)$ to first order in the displacement.

\begin{figure}[t]
\centering
\includegraphics[scale=0.42]{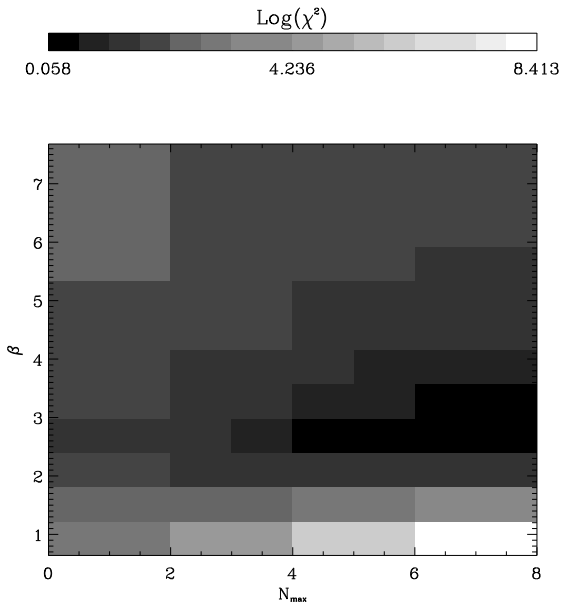}
\figcaption{Dependence of
the shapelet reconstruction on the shapelet parameters $\beta$ and
$N_{\rm max}$. The color scale shows the $\chi^2$ difference
between simulated image parameters and their shapelet reconstruction values, as
defined in Eq. (\ref{eqn:chi2}). Only a small range of $\beta$ and $N_{\rm
max}$ values are acceptable and yield good reconstructions.  For small
$\beta$ values, an increasing $N_{\rm max}$ value quickly result in
poor reconstructions. 
\label{fig:nmax_beta}
}
\end{figure}

The choice of $\beta$ and $N_{\rm max}$ is important to ensure that
each source is faithfully modeled, that the shapelet series converges,
and that our shape measurement is unbiased. To study the first two
issues, we ran a series of simulated grid pointings with the observing
conditions of FIRST, as described in CR02.  The input sources are
described by elliptical Gaussians with major axes, minor axes, and
position angles taken from the FIRST catalog.  Note that this
choice of source model is simplistic and facilitates the convergence
of the shapelet series.  This is however the standard model used in
radio astronomy to parameterize radio sources. In particular, this
parameterization is adopted by the source fitting software used to
generate the FIRST catalog, and is therefore a natural model to use
for our simulations. This choice may affect the exact values of beta
and $N_{\rm max}$ derived, but it will not bias our final shear
estimator as long as these parameters are within the acceptable range
(see Figure~\ref{fig:nmax_beta}). In CR02, we tested our shapelet
reconstruction algorithm in detail and found that it also performs
well for more complicated source models.

With these simulation, we then considered a series of values of
$\beta$ and $N_{\rm max}$ for one of the sources in the pointing,
while keeping these parameters constant for the other sources (all
centroids were also kept fixed).  We also added in noise in the $uv$
plane, and studied the behavior of sources with different SN ratios.
The added noise level is equal to 0.15 mJy per beam in real space,
which is the typical noise level for FIRST.  The lowest SN ratio
considered is 5, which corresponds to the FIRST detection limit; we
explored over two orders of magnitudes of SN ratios, covering the
range appropriate for FIRST sources.  From the reconstructed images,
we find a moderately narrow range of parameters which yield sensible
reconstructions. To quantify this range, we computed the $\chi^{2}$
difference between the input image $f_{\rm in}({\mathbf l})$ and the
fitted image $f_{\rm fit}( {\mathbf l}_{p}; \beta, N_{\rm max})$
defined as
\begin{equation}
\label{eqn:chi2}
\chi^{2} \equiv \frac{\sum_{p=1}^{N_{\rm pix}} 
\left[ f_{\rm fit}( {\mathbf l}_{p}; \beta, N_{\rm max})
- f_{\rm in}({\mathbf l}_{p}) \right]^{2} }{\sigma^{2}_{f{\rm in}} N_{\rm
pix} },
\end{equation}
where the sum is over all $N_{\rm pix}$ pixels within a radius of $5
\beta$ about the centroid.  For convenience, $\chi^2$ was normalized by
the rms $\sigma_{f{\rm in}}$ of the pixel distribution of $f_{\rm
in}$. 

An example of the resulting dependence of $\chi^{2}$ on $\beta$ and
$N_{\rm max}$ is shown in Figure~\ref{fig:nmax_beta}, where the SN
ratio was set to a high value of 500 in this case. The range of
acceptable parameter values is reflected as the region with $\chi^{2}
\sim 1$ on the figure, showing that $\chi^{2}$ is a good measure of
the reconstruction.  In this case, the best reconstruction
is near $(\beta, N_{\rm max})=(1.5,4)$. In the presence of noise,
the best reconstruction tends to favor smaller $N_{\rm max}$ values
while leaving the best $\beta$ unchanged.

We have repeated the above procedure for various sources in different
simulated FIRST grids.  This allowed us to relate the optimal $\beta$
and $N_{\rm max}$ values to the major and minor axes of the sources
derived from the elliptical Gaussian fits listed in the FIRST
catalog. We thus derived an empirical relation between the observed
shape parameters in the FIRST catalog (the deconvolved major and minor
axes) and the optimal shapelets parameters ($\beta$ and $N_{\rm max}$)
for the source ensemble: $\beta=\sqrt{0.9 \times (a/2.35) \times
(b/2.35)}$, and $N_{\rm max}= a/1.5 -1 $, where $a$ and $b$ denote the
FWHM of the major and minor axes. This 'recipe' was then applied to
the real sources in the FIRST survey.

To test the implementation of this recipe in our shear measurement
method, we have performed further simulations. Elliptical Gaussian
sources were first generated with source parameters following the
distribution of the real sources in the FIRST catalog.  These
artificial sources were then sheared with a constant 5\% value for
both $\gamma_{1}$ and $\gamma_{2}$.  The corresponding visibilities
with the FIRST survey settings were generated for over 1,000 simulated
pointings with $\sim 30-40$ sources per pointing.  We then used the
above recipe to determine the shapelet parameters, and applied the
method described in CR02 to calculate the shapelet coefficients for
the $\sim$ 30,000 synthetic sources. The estimators for $\gamma_{1}$
and $\gamma_{2}$ of each source were then calculated according to
Eq.~(\ref{eqn:shear}) applied to different flux-size bins.  We find
that the input shear amplitudes are accurately recovered within the
$1\sigma$ statistical errors. We also performed a null simulation
(with zero shear) and verified that the resulting shear measurement
was consistent with zero.

\subsection{Implementation}
\label{implementation}
Given a set of FIRST flux- and phase-calibrated $uv$ data, sources
positioned within a radius of $64'$ from the phase center are selected
from the FIRST Catalog. This large radius limit ensures bright sources
in the primary beam side lobes are also included in the simultaneous
fit and thus do not contaminate the fainter sources within the primary
beam. Sources which are separated by less than $20''$ are merged into
one source, as simulations show that treating close companions as one
source yields the best shapelet fits (see also the discussion in
\S\ref{fragment}). The shapelet parameters are determined according to
the prescription described in \S\ref{params}. We then compute the
source shapelet coefficients and their covariance matrix using a
least-squares fit to the visibilities, where all sources in a given
pointing are fitted simultaneously.  In principle, some of the
systematics that distort the shape of sources could be corrected for
at this stage. However, to save computing time, we decided to correct
for the systematics afterwards, as described in the next section.  The
whole FIRST data set observed up to year 2001 were processed on the UK
Cosmos Supercomputer. The total of about 50,000 individual pointings
required about 9,000 hours of Cosmos CPU time.

As a first check, the output coefficients were examined by computing
the source fluxes and centroids derived from the shapelet coefficients
(see CR02).  Along with those fields flagged when the fit failed to
converge, we have discarded about $3\%$ of the processed sources, due
to corrupted observational data, numerical convergence problems and,
in a few cases, inadequate choice of input shapelet parameters.  On
average, the computed shapelet flux densities and centroids agree
rather well with those in the FIRST catalog, which used an elliptical
Gaussian fit to measure source intensities and positions.  The
shapelet flux densities and flux density errors were then combined to
compute the signal-to-noise ratio of each source. To control
systematics, we discarded measurements with observing parameters $|HA|
> 4$ hours and $\sqrt{l^2+m^2} > 20'$, and only kept sources with an
integrated flux density $\ge$ 1mJy and deconvolved major axis $< 7''$.
Excluding the measurements for point sources (deconvolved major axis
$< 2''$), we thus were left with $\sim 3.6 \times 10^5$ usable shear
estimators with associated measurement errors.

Since the FIRST pointings partially overlap (see Becker et al. 1995
for details), each source is observed from one to four times.  We
verified that the corrected shear estimators of a given source in
different pointings are consistent within the errors.  The shear
estimators of a given source are then coadded using
the square of the observed signal-to-noise ratio as a weight. The sky
coverage in the northern cap is close to 8000 $\deg^{2}$, and the
resulting resolved (major axis $>2''$ and $<7''$) source number
density is about 20 sources $\deg^{-2}$.

\section{Summary of Systematic Effects}
\label{systematics}

Because the weak-lensing signal is only on the order of 1\%,
systematic effects must be carefully accounted for, as they may
otherwise introduce spurious shear correlations. The dominant
systematics arise from the anisotropy of the synthesized 
beam, which is directly related to the $uv$ sampling, and thus on the
geometry of the interferometric array. Note that, unlike the case
with optical observations, systematic effects in our case are well-known and
deterministic (albeit somewhat complicated) and can be calculated 
to high-degree of accuracy. In the following, we briefly
describe the different sources of systematic effects and show that
they are best viewed as functions of four observing parameters (see
e.g., Perley et al. 1989; Taylor et al. 1999; \& Thompson et al. 1986 for a detailed
description of these effects).

\subsection{The HA-DEC Effect}
\label{hadec}
Interferometric data are collected in Fourier space, or $uv$ space, at
a finite and discrete number of points. The Fourier-transform of the
sky surface brightness is thus multiplied by the $uv$ sampling
function which determines the shape and size of the convolution beam
(PSF) in real-space. The $uv$ sampling function is deterministic and
is a sum of delta functions centered on the positions
\begin{eqnarray}
\label{eqn:uvw}
\nonumber
\left( \begin{array}{c} u \\ v \\ w \end{array} \right) & = & {1 \over \lambda}
\left( \begin{array}{ccc} \sin H & \cos H &
0 \\ - \sin \delta \cos H & \sin \delta \sin H & \cos
\delta \\ \cos \delta \cos H & - \cos \delta \sin
H & \sin \delta \\ \end{array} \right) \left( \begin{array}{c}
 L_x \\ L_y \\ L_z \end{array} \right), \\
\end{eqnarray}
where $\lambda$ is the observing wavelength, $(L_x, L_y, L_z)$ are the
VLA antenna spacings, and $H$ and $\delta$ are the source hour-angle
and declination. Since the former two are fixed and known for a
given observation, the effective beam can thus be calculated exactly
from the source position on the sky. This can be seen as the
projection of the antenna plane onto the $uv$ plane, which depends on
the source position.

\begin{figure}[t]
\centering
\includegraphics[scale=0.42]{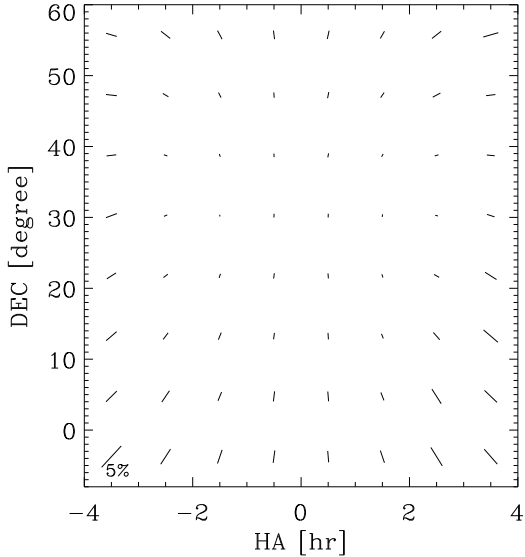}
\figcaption{The
artificial shear pattern induced by the HA-DEC effect.  The length and
direction of the plotted lines indicate the amplitude and orientation
of the distortion, respectively.  The line at the bottom-left
indicates a distortion of 5\%.
\label{fig:hadec}
}
\end{figure}

With the knowledge of the $uv$ sampling function, we place a point
source at the phase-tracking center and generate simulated
visibilities using the FIRST observing settings.  The simulated
visibilities are modeled using the Shapelet method, and the artificial
shear of the simulated source due to systematics are calculated using
the fitted Shapelet coefficients.  We then repeat the procedure for
various values of $(H, \delta)$. The resulting shape distortion is
shown in Fig \ref{fig:hadec}. The distortion is minimal at small hour
angles and at declination near 34 degrees, corresponding to the
latitude of the VLA, when sources are directly overhead. The
induced distortion is $\sim 5\%$ at $(H,\delta) =
(3.5^{hr},-5^{\circ})$, roughly in the radial direction with respect
to $(H,\delta) = (0,34^{\circ})$.

\subsection{The Non-coplanar Effect}

The VLA is a two-dimensional array laid out across the surface of
the Earth.  Due to the curvature and rotation of the Earth, the
visibilities are not strictly measured in a plane but in the
three-dimensional Fourier space labeled as $(u,v,w)$:
\begin{eqnarray}
\nonumber
\label{eq:non_coplanar}
V(u,v,w) &=& \int \int {dl dm \over \sqrt{1-l^2-m^2}} I(l,m) \times \\ 
&&e^{-2 \pi i[ul+vm+w(\sqrt{1-l^2-m^2}-1)]},
\end{eqnarray}
where for simplicity we have ignored other systematic effects which we
describe below. For short observing durations and for sources
close to the phase tracking center, the $w$ term is small and is often
dropped. In this approximation the visibility function reduces to a
two-dimensional Fourier transform
\begin{equation}
V(u,v) = \int \int {dl dm \over \sqrt{1-l^2-m^2}} I(l,m) e^{-2 \pi i[ul+vm]},
\end{equation}
which has the advantage of allowing fast computation.  This
approximation induces a distortion of the source images that depends
on the amplitude of $|w(\sqrt{1-l^2-m^2}-1)|$, i.e., on the distance
from the phase tracking center $(l=0,m=0)$. As mentioned in the
previous section, the amplitude of $w$ depends on the source observing
hour angle and declination.  Thus, the shape distortion due to the
two-dimensional approximation is a function of the source position
$(l,m)$ with respect to the phase center, as well as the observing
hour angle and declination.  In our Shapelet approach, we use the
two-dimensional approximation in order to save computing time, and
therefore the distortion must be corrected for.

\begin{figure}[t]
\centering
\includegraphics[scale=0.42]{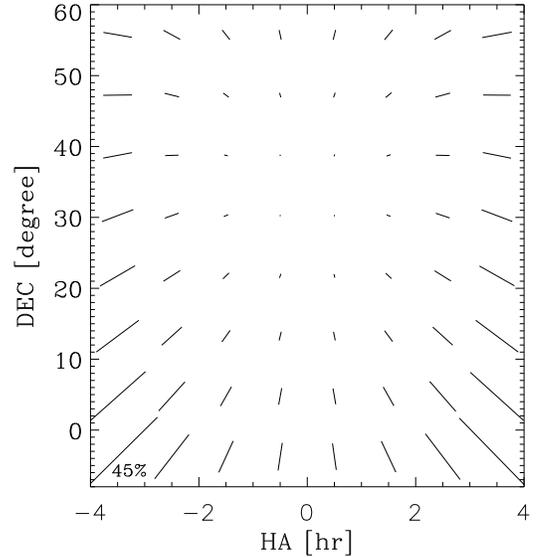}
\figcaption{The artificial shear pattern induced by the non-coplanar effect,
for the case of $(l,m)=(7',7')$ away from the phase-tracking center.
The line at the bottom-left indicates a distortion of 45\%.
\label{fig:noncoplanar}
}
\end{figure}

We quantify this distortion using simulations similar to that
described in \S\ref{hadec}, but with sources placed at various
off-center positions in the simulated grid using
equation~(\ref{eq:non_coplanar}). Fig. \ref{fig:noncoplanar} shows an
example of the resulting distortion pattern, for which the simulated
source is placed at ($l$,$m$)=$(7',7')$. In this setting, the
distortion at $(H,\delta) = (0^{hr},-5^{\circ})$ is $\sim 20\%$, and
is roughly in the radial direction with respect to$(H,\delta) =
(0,34^{\circ})$. This effect is the dominant source of systematic
effects (see \S\ref{simulations}).

\subsection{Bandwidth Smearing}
The bandwidth-smearing effect with interferometers is exactly analogous
to chromatic aberration in an optical telescope.  It is due to the fact
that radio interferometric observations are not monochromatic, but are
instead done within a finite frequency band. As a result, the
visibility registered at a particular $uv$ position is actually an
average of the actual visibilities over a small interval of $(u,v)$
positions extended in the radial direction.  The delay tracking, which
corrects for the difference in arrival time of light rays at two
antennas prior to correlation of the signals, is only accurate at the center of the
field-of-view and at the central observing frequency.  Suppose that the observing
central frequency is $\nu_0$ and the phase tracking center is $(l_0, m_0)$.
Signals at frequency $\nu$ arriving from sky position $(l, m)$ will
have a phase delay error of $(u_0 l + v_0 m)/ \nu_0$, resulting in 
a phase shift of $2 \pi (\nu - \nu_0) (u_0 l + v_0 m)/ \nu_0$.
The smeared visibilities, $\tilde{V}$, are thus (Perley et al. 1994)
\begin{eqnarray}
\nonumber \tilde{V}(u_0, v_0)&=&{1 \over \int d\nu' G(\nu')} \int
d\nu' V \left( u_0 {\nu \over \nu_0}, v_0 {\nu \over \nu_0} \right)
({\nu \over \nu_0})^2 \\ & & \times ~G(\nu') e^{2 \pi i {\nu' \over
\nu_0} (u_0 l + v_0 m)},
\end{eqnarray}
where $G(\nu)$ is the passband function and $\nu' = \nu - \nu_0$.  For
simplicity, here and in the following, we ignore the effect of the
primary beam power pattern and the weighting function in the $uv$
plane. In real space, this corresponds to a convolution of the true
sky intensity, $I(l,m)$, with a distortion function $D$, such as
$\tilde{I}(l,m) = I(l,m) \ast D(l,m,\nu)$, where $D$ is defined as
\begin{eqnarray}
D(l,m,\nu')= \int \int du_0 dv_0 e^{2 \pi i ( u_0 l + v_o m)}
\times~~~~~~~ \nonumber \\ \left[{1 \over \int d\nu' G(\nu')} \int
d\nu' G(\nu') e^{2 \pi i {\nu' \over \nu_0} (u_0 l + v_0 m)} \right],
\end{eqnarray}
provided that the fractional bandwidth is sufficiently small.  The
amplitude of bandwidth-smearing distortion therefore depends on the
source position (with respect to the phase tracking center), and on
the size of the frequency interval.  Since a change in frequency moves
the $uv$ points radially on the $uv$ plane, bandwidth
smearing distorts the observed images in the radial direction with
respect to the field center.

\begin{figure}[t]
\centering
\includegraphics[scale=0.42]{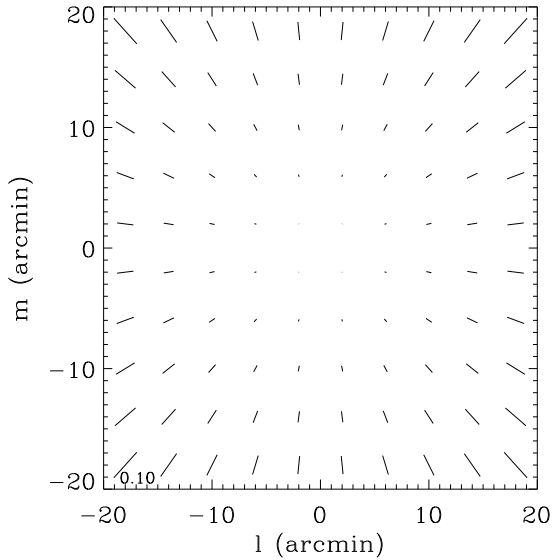}
\figcaption{The artificial shear pattern induced by the bandwidth
smearing effect.  The length and direction of the plotted lines
indicate the amplitude and orientation of the distortion,
respectively. 
\label{fig:bandwid}
}
\end{figure}

For FIRST, $G(\nu)$ can be approximated by a square passband with a
width of 3 MHz.  The distortion induced $20'$ away from the center of
the field-of-view is about 6\%.  The resulting shear pattern is shown
in Figure~\ref{fig:bandwid}.

\subsection{Time Averaging}
Time-averaging smearing is another systematic effect that results
from data averaging. Correlated signals received by pairs of antennas
are, in practice, averaged over small time intervals in order to reduce
the size of the visibility files.  During a time-interval $\delta t$,
the Earth rotates by an angle of $\omega_e \delta t$, where $\omega_e$
is the Earth angular rotation velocity. For observations at
a declination of $90^{\circ}$, the Earth rotation corresponds to a
tangential rotation in the $uv$ plane.  The true visibilities are
therefore averaged over an interval of $uv$ positions, extended
tangentially, which results in a tangential distortion of images in
real space.  The time-averaging effect at declination $90^{\circ}$
thus bears an interesting analogy to the bandwidth smearing effect,
which distorts images in the radial direction.

In general, however, the time averaging effect is more complex.
Depending on the source position relative to the antenna (Earth)
space, the loci on the $uv$ plane over which visibilities are being
averaged varies in length and shape.  A convenient way to estimate the
size of image distortion is to calculate the response of a point
source. Since the time averaging effect preserves the integrated flux
density, the induced tangential broadening must be compensated for by
the reduction in the peak amplitude of the point source response.

Averaging a waveform of frequency $\nu$ over a time interval $\delta
t$ reduces the amplitude of the response by sinc$(\nu \delta t)
\approx 1-(\pi \nu \delta t)^2 / 6$, for $\nu \delta t \ll 1$, where
$\nu \delta t$ corresponds to the phase change.  For a source at
$(l,m)$ with respect to the phase-tracking center, the instantaneous
phase rate is $({du \over dt}l + {dv \over dt}m)$, and therefore,
integrating over a small time interval $\delta t$, the reduction of the
amplitude of the point source response is
\begin{equation}
R_{\delta t} = {I \over I_0} \approx 1-{\pi^2 {\delta t}^2 \over 6}
\left( {du \over dt} l + {dv \over dt} m \right)^2, 
\end{equation}
where
\begin{eqnarray}
{du \over dt} &=& {\omega_e \over \lambda} ( L_x \cos H - L_y \sin H)
\nonumber \\
{dv \over dt} &=& {\omega_e \over \lambda} ( L_x \sin \delta \sin H +
L_y \sin \delta \cos H),
\end{eqnarray}   
which follows directly from Equation~(\ref{eqn:uvw}), and $\omega_e =
{dH \over dt}$.

Combining the above two equations, we can estimate the reduction of
response for sources at any given observing coordinates $(H,\delta)$.  The
amplitude of the tangential width broadening due to time-averaging
smearing is then simply $R_{\delta t}^{-1}$.  The distortion is
therefore a function of the antenna configuration ($L_x$ and $L_y$), 
the source position in the sky (the hour angle and declination), and 
the source position with respect to the center of the field-of-view 
$(l,m)$.

\begin{figure}[t]
\centering
\includegraphics[scale=0.42]{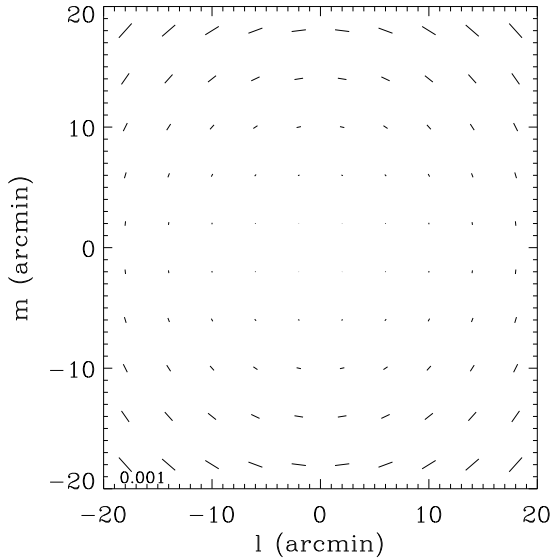} 
\figcaption{The artificial shear pattern induced by the time-averaging
smearing effect for the case of $\delta=90^{\circ}$.  The distortion
pattern is tangential in this case.  The pattern varies with sky
position, but in general, the distortion is on the order of 0.1\% at
the edge of the field-of-view of concern.
\label{fig:timeave}
}
\end{figure}

For FIRST, the averaging time-interval is five seconds. The resulting
distortion pattern for $\delta=90^{\circ}$ in shown in
Figure~\ref{fig:timeave}.  The distortion from time averaging smearing
is small in the FIRST settings; for most sources the
distortion is less than 0.1\%.

\subsection{Primary Beam Attenuation}
\label{pbeam}
In practice, the true sky surface brightness $I(l,m)$ is multiplied by
the antenna power pattern of the primary single dish $A(l,m)$ before
signals are  correlated.  The primary beam power pattern is
closely related to the diffraction pattern of a circular dish, and can
be approximated by (Condon et al. 1998)
\begin{equation}
\label{eqn:pbeam}
A(r)= \left({2 J_1({r \over r_0}) \over r/r_0}\right)^2,
\end{equation}
where $J_1$ is the Bessel $J$ function of order 1, $r \simeq 3.23
\times \sqrt{l^2+m^2}$, and $r_0$ is the FWHM of the primary beam
pattern.  For FIRST, $r_0=30'.83$.

\begin{figure}[t]
\centering
\includegraphics[scale=0.42]{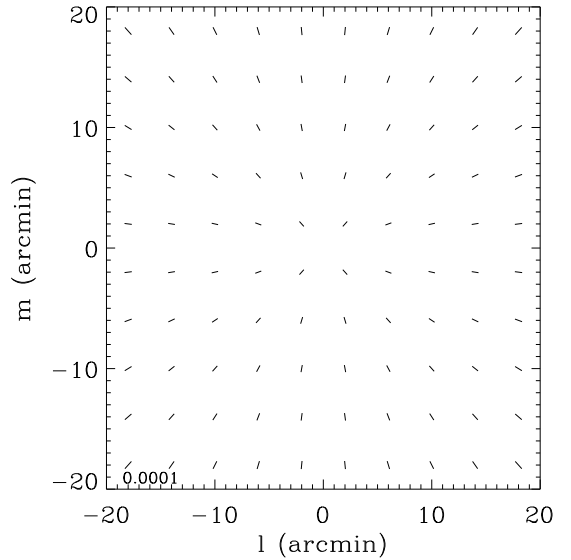}
\figcaption{The
artificial shear pattern induced by the primary beam power pattern
attenuation.  The resulting distortion is radial and very small.
For a big circular Gaussian source with a FWHM of $10''$, the distortion
is less than 0.01\%.
\label{fig:pbeam}
}
\end{figure}

The primary beam pattern introduces a variation in sensitivity in the
radial direction across the observing field-of-view which modifies the
observed shape of source images.  A general treatment of this effect
is detailed in the Appendix.  The effect depends on the primary beam
power pattern as well as on the source sizes; for larger sources the
distortion is more severe. Using Equations (\ref{eqn:pbeam}) and
(\ref{eqn:epbeam}), we calculate the effect for typical FIRST sources.
The distortion is in the radial direction with respect to the field
center, and the induced shear pattern for a circular Gaussian source
with FWHM of $10''$ ($\sim 2$ times the beam size) is shown in
Fig~\ref{fig:pbeam}.  The distortion is less than 0.01\% and thus the
effect can be safely ignored in the FIRST case.

\subsection{Source Fragmentation} 
\label{fragment}
A significant fraction of radio sources have a double-lobe structure,
and are often broken into two components by the FIRST object finder.
Because these radio lobes tend to be aligned, this produces strong
shear correlations on small angular scales.  As mentioned in
\S\ref{implementation}, we have treated sources within $20''$ of each
other, the typical separation of double-lobe sources, as a single
source for the shapelet fitting. Furthermore, for sources that lie
within $1'$, only one source is selected at random for use in the
shear measurements. This leads to a loss of information only on scales
smaller than $1'$, a scale much smaller than the scales on which we
measure the weak lensing signal (greater than about $30'$; see
\S\ref{results} below)

\begin{figure*}[ht]
\centering
\includegraphics[scale=0.42]{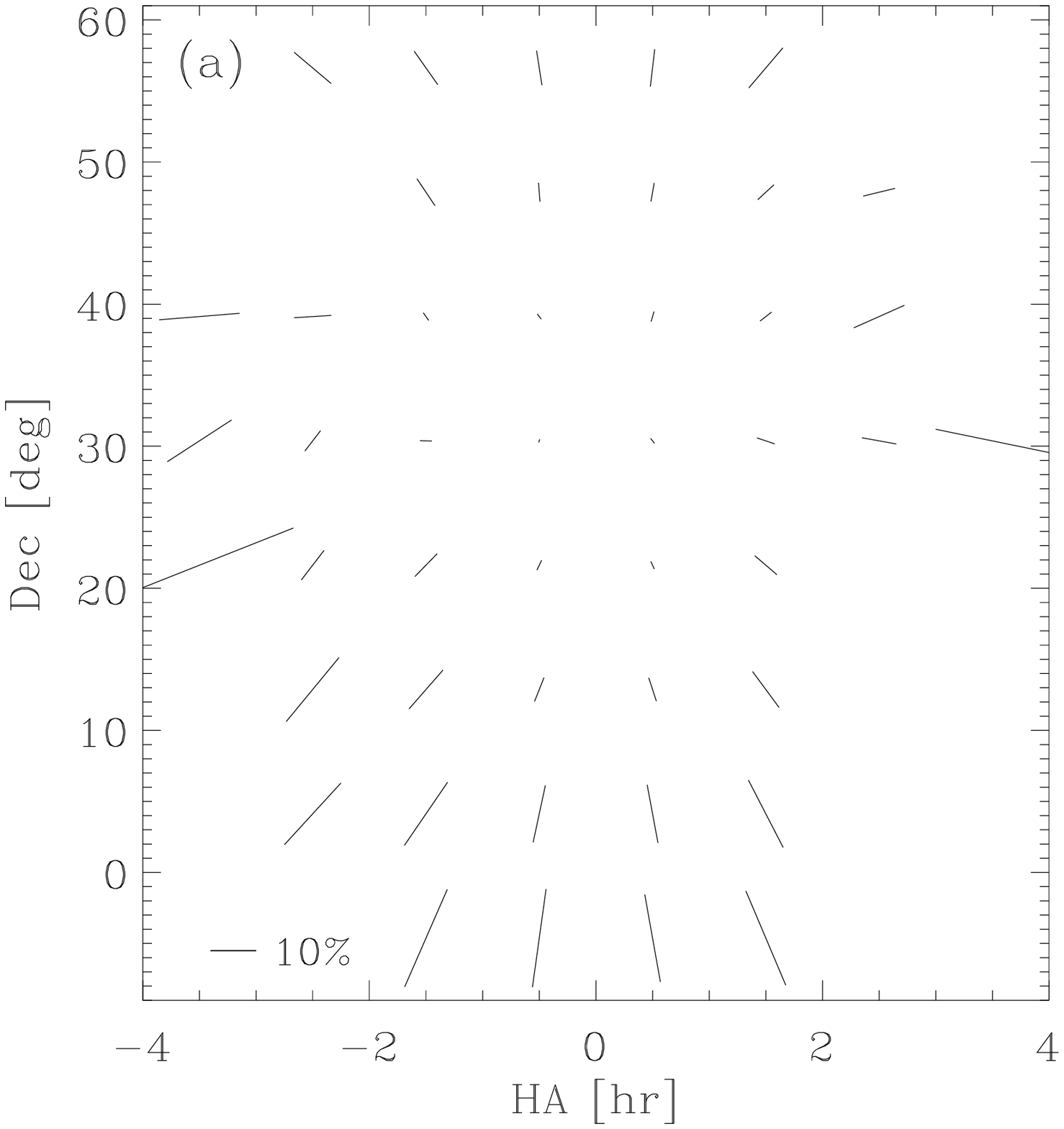}
\includegraphics[scale=0.42]{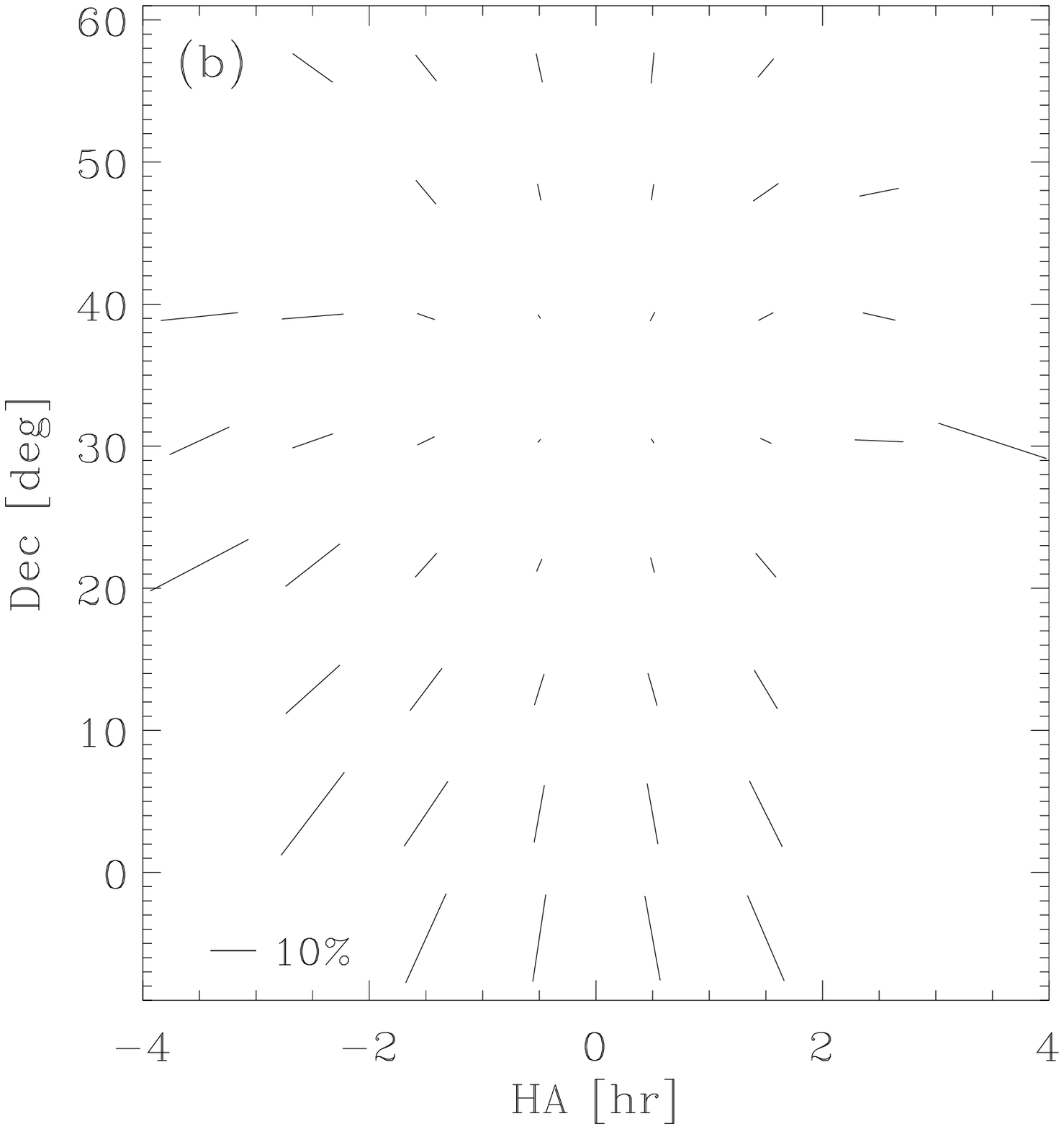}
\figcaption{(a) The shear pattern in the HA-DEC plane from the averaged
shear measurements.  The line at the bottom left corner indicates an
amplitude of 10\% shear. (b) The shear pattern in the HA-DEC plane from the simulated
shears due to the systematic effects.  The line at the bottom left
corner indicates an amplitude of 10\% shear.
\label{fig:data_hadec}
}
\end{figure*}

\begin{figure*}[ht]
\centering
\includegraphics[scale=0.42]{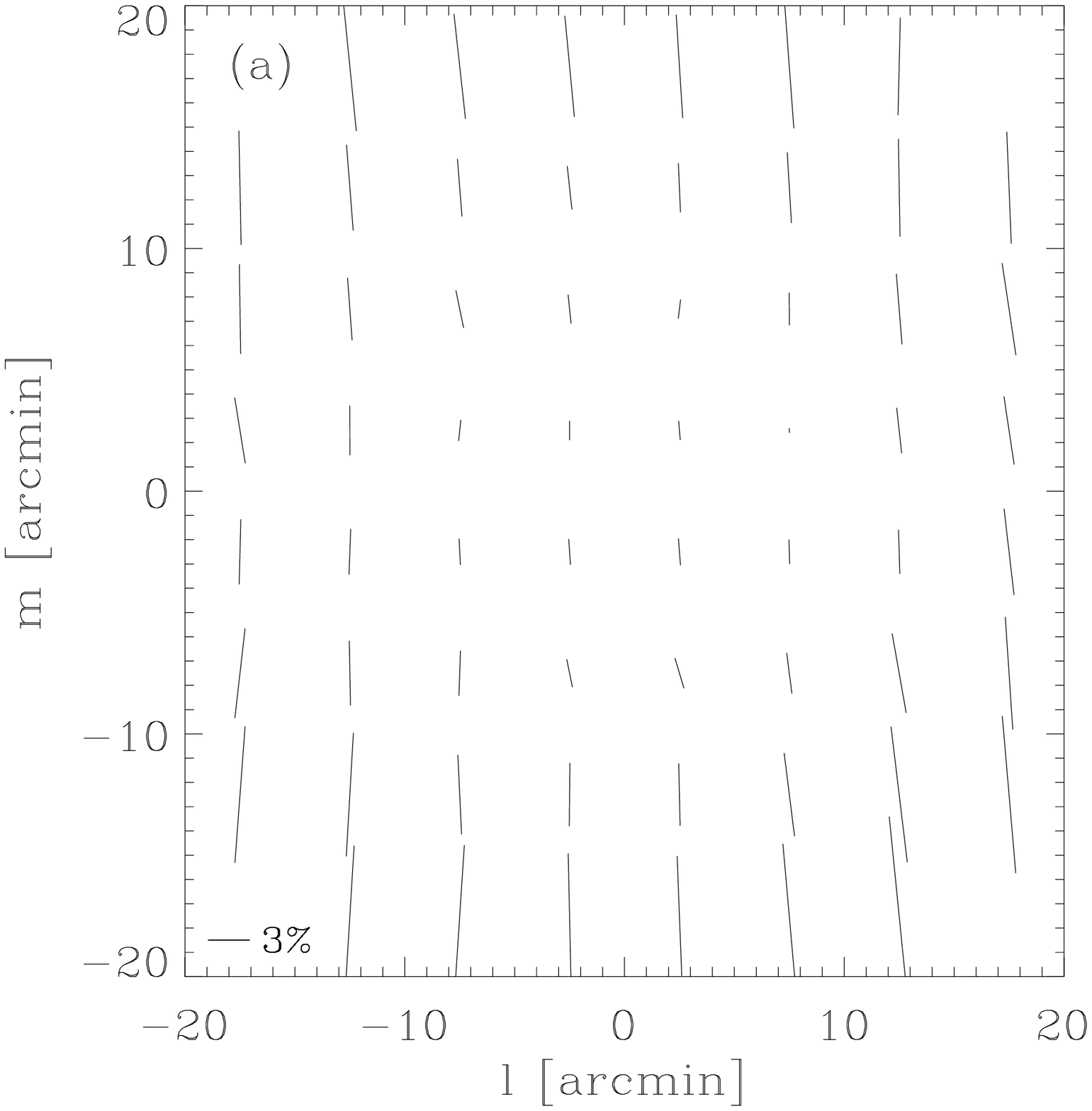}
\includegraphics[scale=0.42]{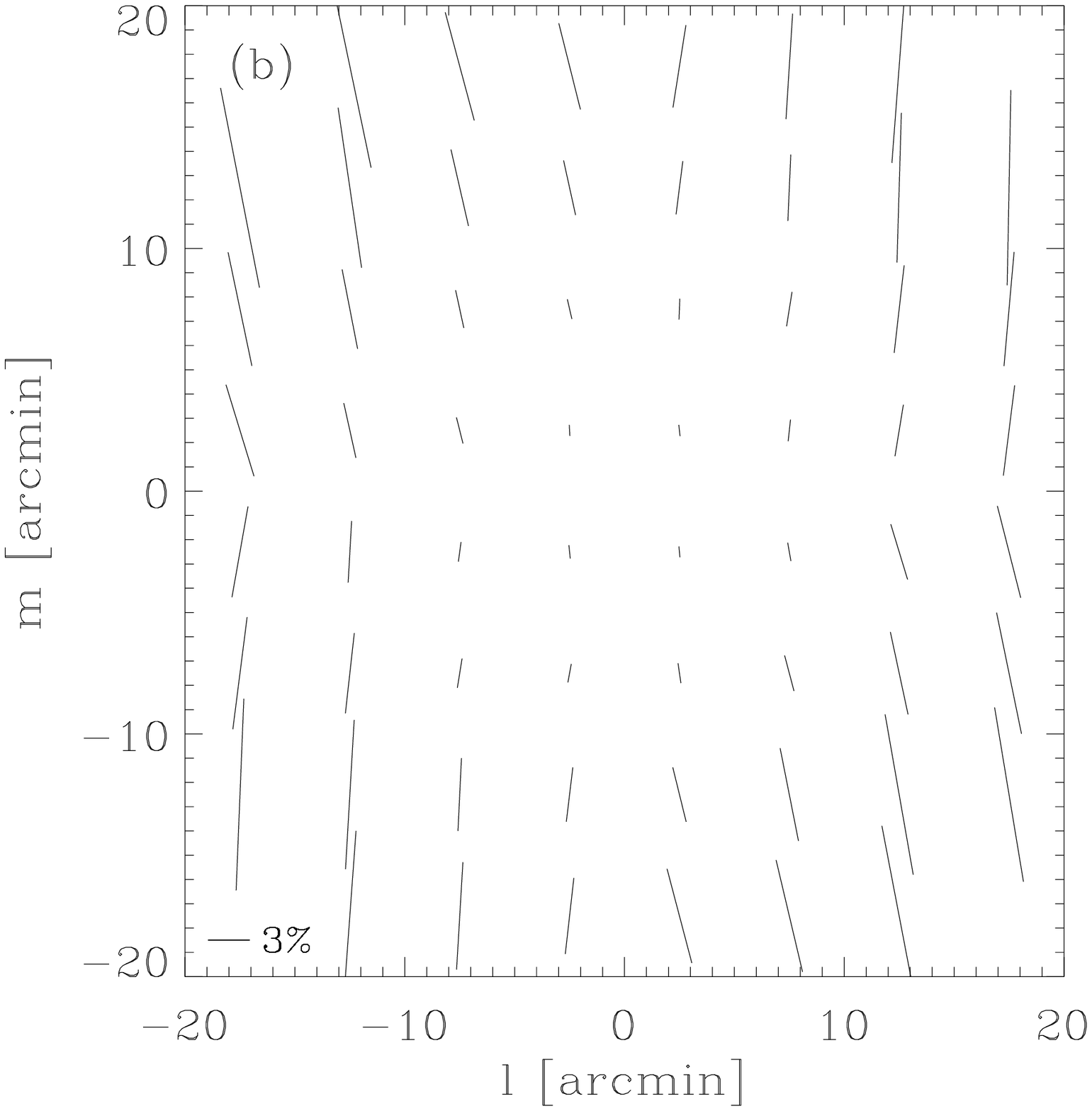}
\figcaption{(a) The shear pattern in the $l$-$m$ plane from the averaged shear
measurements.  The line at the bottom left corner indicates an
amplitude of 3\% shear. (b) The
shear pattern in the l-m plane from the simulated shears due to the
systematic effects. The line at the bottom left corner indicates an
amplitude of 3\% shear.
\label{fig:data_xy}
}
\end{figure*}

\section{Corrections for Systematics}
\label{sys_corrections}

\subsection{Observations}
Without correction, the shear measurements of our FIRST sample are
dominated by the systematic effects described above. These
effects are complicated and are correlated with each other, and can
not easily be decomposed into convolutions and distortions, as is
the case with optical observations. To correct them, we therefore
adopt a statistical approach by measuring and subtracting the systematic
shear as a function of the observation parameters (RA, DEC, l, m)
and of the source size. As a test, we also apply the same procedure to
the simulations.

Figs. \ref{fig:data_hadec}a and \ref{fig:data_xy}a show the average of
the shear estimators for each of the $\sim 5.3 \times 10^{5}$
measurements (with major axis $>2''$ and $<17''$ before coaddition) in
HA-DEC and l-m bins for all source sizes. Since the lensing signal
averages out in these cells (see \S\ref{corrections} for more
details), the resulting patterns give a measure of the systematics.

The HA-DEC plane shows a large-scale pattern in $15 \times 6$ degree
bins. Most (75\%) of the measurements concentrate in the central two
HA columns, $|HA| < 1$, which have the highest statistical accuracy.
The shear pattern in the l-m plane is less prominent, and shows a
predominantly vertical pattern.  The measurements are more evenly
distributed in this plane, although the number density decreases at
larger radii.  Note that all measurements with $|HA| > 4$hrs and
$\sqrt{l^2+m^2} > 20'$ have been excluded at an earlier stage. The
systematics shown in these two planes are coupled: the minimum shear
distortion occurs at around $HA=0$, Dec~$\simeq 34^{\circ}$, where
sources are directly overhead from the VLA, and at the phase tracking
center $l=0, m=0$.  The farther away from these two central values, the
greater the spurious shear due to systematics.  We also examined the
dependence of the systematics on the source size cuts, and found that
smaller sources are subject to larger systematic distortions, as
expected. We do not find any significant dependence on observation
epoch over the eight years of the observing campaign.

\begin{figure*}[ht]
\centering
\includegraphics[scale=0.42]{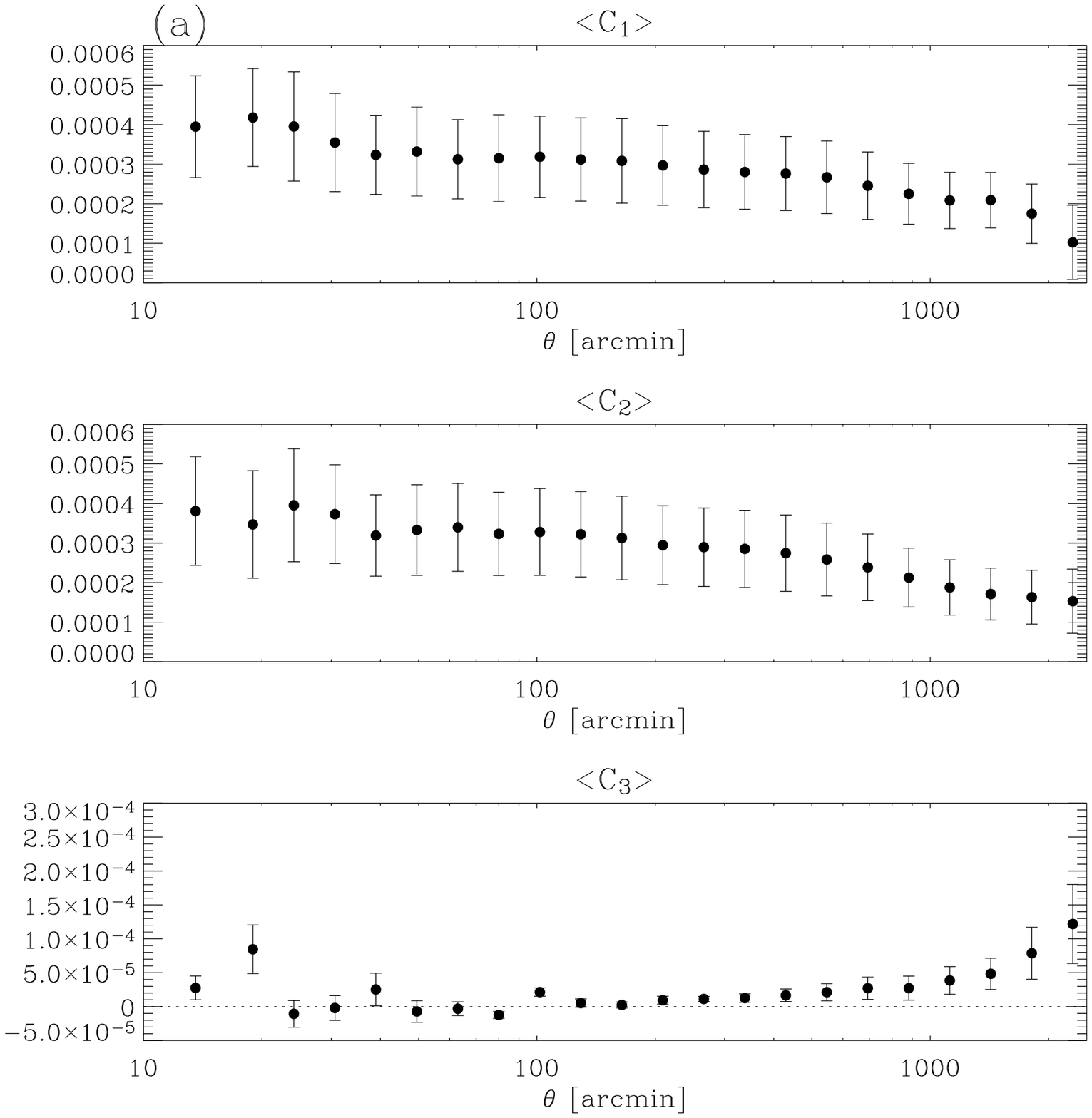}
\includegraphics[scale=0.42]{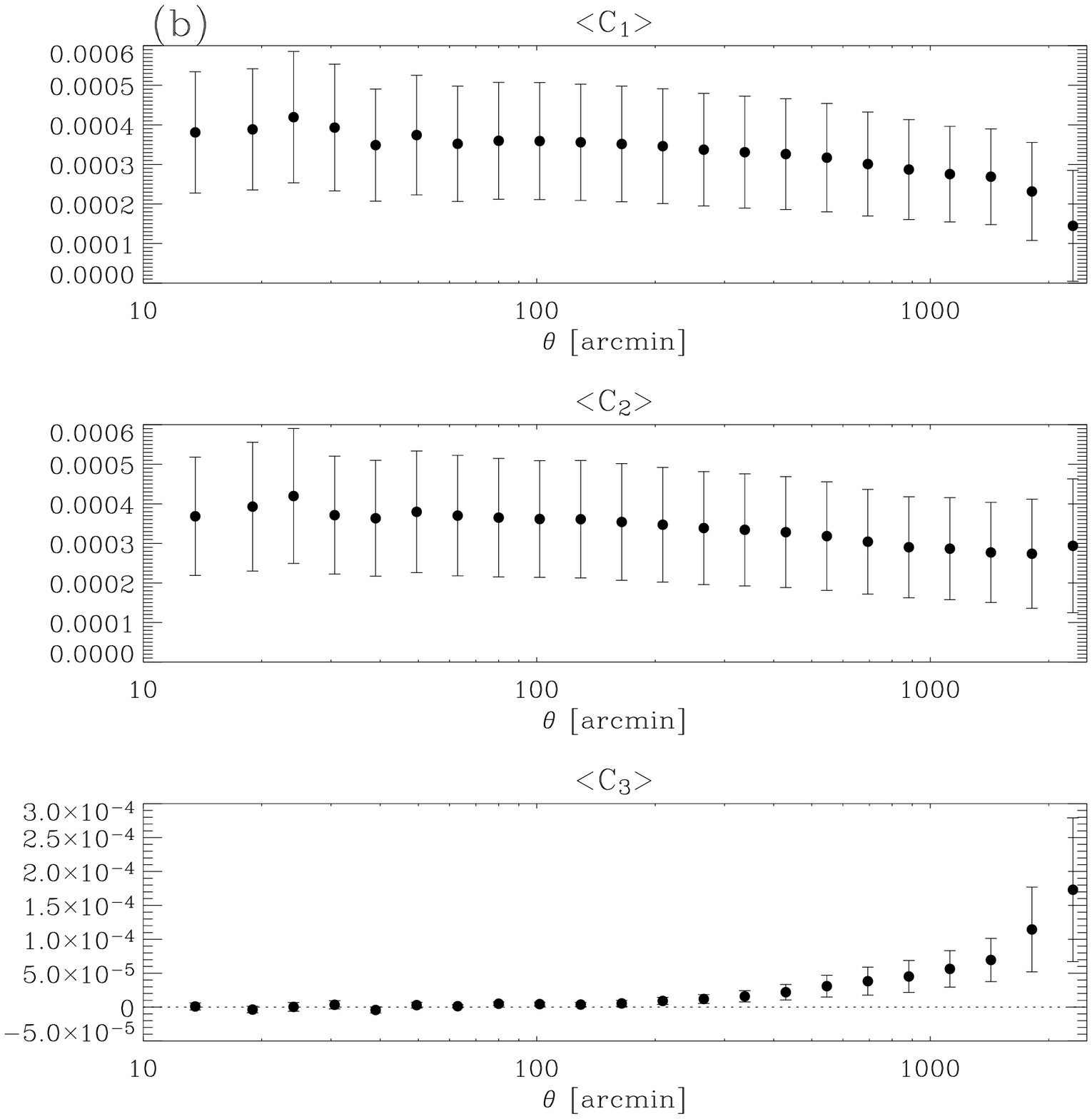}
\figcaption{
(a) The shear correlation function of the data without
correction for the systematics. The error bars are computed from
field-to-field variations, and include statistical errors and cosmic
variance. (b) The shear correlation function of the simulations without
corrections for the systematics. The 1-$\sigma$ error bars are
computed from field-to-field variations.
\label{fig:data_correlation}
}
\end{figure*}

\subsection{Simulations}
\label{simulations}
As discussed in \S\ref{systematics}, the largest systematic
distortions for FIRST are the coupled HA-DEC and non-coplanar effects,
which produce as much as $\sim$ 50\% artificial shear when a 
source combines low declination, with an observation at large 
hour angle, and at a position far from the phase
tracking center.  Another important contribution is from the
bandwidth-smearing effect, which produces $\sim$ 6\% shear $20'$
from the phase tracking center.  Although the systematic effects
are well-understood and calculable to high accuracy, their impact on
the shear statistics is complicated, since source observations do not
uniformly occupy the HA-DEC-l-m space and since we apply complicated
cuts during the analysis.  

Ideally, we could use a control sample, such as point sources, to
quantify the systematic effects and check our correction method. Since
point sources carry no lensing information, any measured shear must
indeed be due to systematic effects.  However, in our shapelets
approach, the shear estimators require fitting coefficients up to
$n=4$ (Eq. \ref{eqn:shear}), which is a poor choice of input
parameter for point sources (see, e.g., Fig. \ref{fig:nmax_beta}).
This not only compromises the fits for point sources, but also affects
the reconstruction of resolved sources, since all sources in a given
field are fitted simultaneously.  Although we could imagine increasing
the size of $\beta$ for point sources to yield a more reasonable fit,
this would be difficult to implement given the presence of other
sources (leaving aside the problem of quantifying
the effects on these sources).  In
addition, this would have meant doubling the fitting parameters in the
model and thus doubling the computing time, which was already
substantial.  More importantly, the systematic effects can in fact be
understood and calculated very accurately, as we will see below. The
averaged data described in the previous section also serves as another
check for systematics.  To quantify the systematics at the required
precision, we therefore have carried out further simulations to 
study the impact on the shear statistics.

For this purpose, we consider the HA-DEC, non-coplanar, and
bandwidth-smearing effects and ignore the other effects, which are at
least one order of magnitude smaller in amplitude, or have been
incorporated directly in the analysis (such as the source
fragmentation effect). Using the FIRST Catalog, we begin by
randomizing the source position angles, which, along with the major
and minor axis information, gives a measure of the intrinsic
ellipticity distribution. For every source, we computed the combined
systematic distortions listed above for each individual observation;
the result is subsequently convolved with the source's intrinsic shape and
multiple observations are coadded.  To avoid large computing time, we calculated the
systematics on a grid in HA-DEC-l-m space, and interpolate between the
grid positions. Examples are shown in Figs. \ref{fig:data_hadec}b and
\ref{fig:data_xy}b, where we average the simulated measurements in the
same HA-DEC and l-m bins as for the observations. The resulting shear
patterns and amplitudes are almost identical to those of the data (see
Figs. \ref{fig:data_hadec}a and \ref{fig:data_xy}a excepting for the l-m patterns
which show small discrepancies at the edges of the field where the data are
of low statistical significance.  We are therefore confident that the
systematics estimation is, at least on average, a good representation
of the true systematics for individual sources.

\begin{figure*}[ht]
\centering
\includegraphics[scale=0.42]{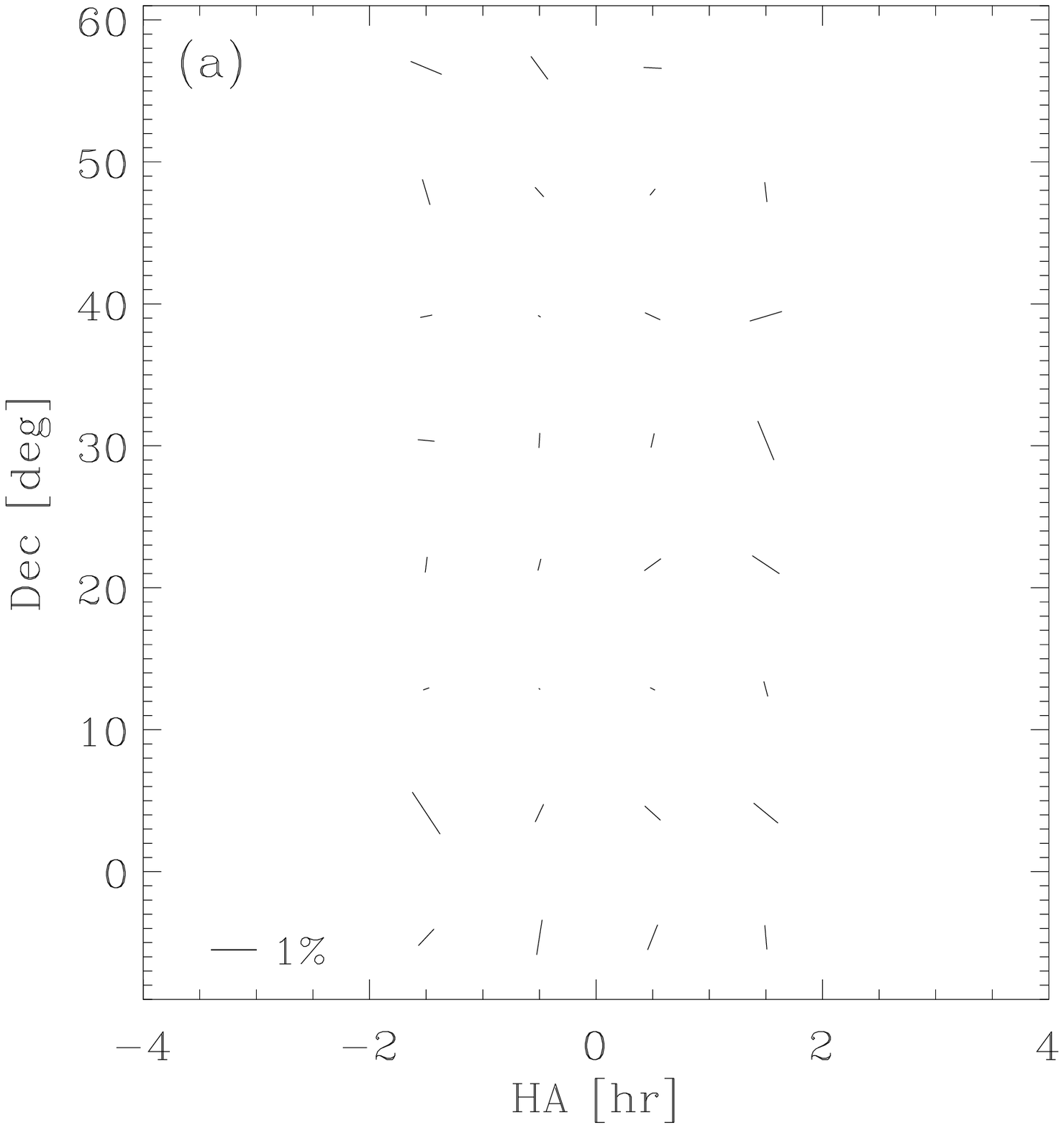}
\includegraphics[scale=0.42]{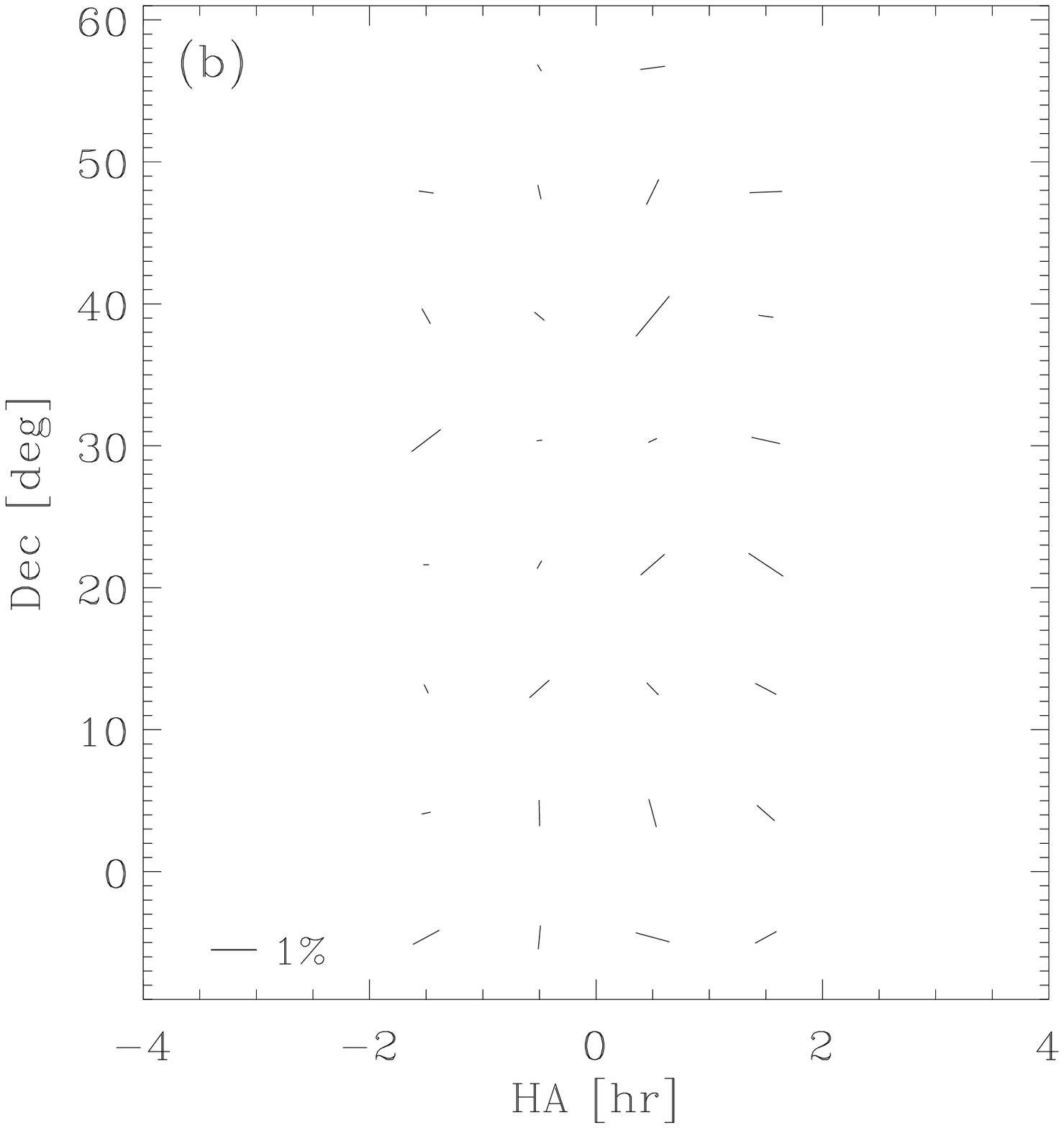}
\figcaption{(a) The shear pattern from
    the averaged shear measurements after the HA-DEC corrections.  The
    line at the bottom left corner indicates an amplitude of 1\%
    shear. (b) The shear pattern from the simulated shears after the
    HA-DEC corrections.  The line at the bottom left corner indicates
    an amplitude of 1\% shear.
\label{fig:data_hadec_correct}
}
\end{figure*}

\subsection{Uncorrected Shear Correlation Functions}
\label{correlation}

We compute the rotated shear correlation functions, $C_1$, $C_2$, and
$C_3$, as described in \S\ref{formalism}, for both the simulation and the
data. Operationally, the shear correlation functions are measured as
\begin{eqnarray}
\nonumber C_1(\theta)&=&{\sum_{i,j}w_i w_j \gamma_t({\bf \theta}_i)
\gamma_t({\bf \theta}_j) \over \sum_{i,j}w_i w_j} \\
C_2(\theta)&=&{\sum_{i,j}w_i w_j \gamma_r({\bf \theta}_i)
\gamma_r({\bf \theta}_j) \over \sum_{i,j}w_i w_j}
\end{eqnarray}
where the sums are over all pairs of galaxies i and j, and $\theta =
|{\bf \theta}_i - {\bf \theta}_j|$ is the separation of pairs within a
bin. Similarly, for $C_3$, the $\gamma$'s are replaced by
$\gamma_t({\bf \theta}_i)$ and $\gamma_r({\bf \theta}_j)$,
respectively.  We adopt the optimal weight function derived by
Bernstein \& Jarvis (2002)
\begin{equation}
w={1 \over \sqrt{\gamma^2 + (1.5\sigma_{\eta})^2}},
\end{equation}
where $\sigma_{\eta}$ is the measurement uncertainty in each shear
component, as measured by the Shapelet method, and transformed to the
sheared coordinates in which the shape is circular.  We divide the
data into twelve subsamples according to their sky position, and
compute the correlation functions separately.  The correlation
functions are calculated in bins of $\Delta\ln(\theta)=0.2$, and the
1-$\sigma$ error bars are quantified using the field-to-field
variation of the twelve subsamples.  The error bars include
noise from uncertainties in the shape measurements, from cosmic
variance, and from intrinsic dispersion of the (unlensed) source
shapes, which by averaging over the entire sample is measured to be
around $\sim 0.4$.
 
The data shear correlation functions, without any corrections, are
heavily dominated by the systematics, and we expect their correlation
functions to have high amplitude and to appear similar to those of the
simulation. As demonstrated in Figs. \ref{fig:data_correlation}a and
\ref{fig:data_correlation}b, the general behaviors of the $C_{1}$ and
$C_{2}$ functions for the data and of the simulation are in very good
agreement from $10'$ to $40^{\circ}$. For the $C_{3}$ function, the
simulation captures the large-scale behavior well but does less well
on small angular scales.  It is nevertheless reassuring that our
knowledge of the systematics is realistic.

Note that the measurements of $C_1$ and $C_2$ on different angular
scales in Fig.\ref{fig:data_correlation} appear correlated both for
the data and for the simulations. This reflects the fact that the
systematics, which dominate $C_1$ and $C_2$ before the corrections,
are strongly correlated within a patch of the sky used to measure the
correlation functions. On the other hand, $C_3$ (for $\theta \lesssim
1000'$) is less correlated, and therefore its errors are dominated
by statistical uncertainties.

\begin{figure*}[ht]
\centering
\includegraphics[scale=0.42]{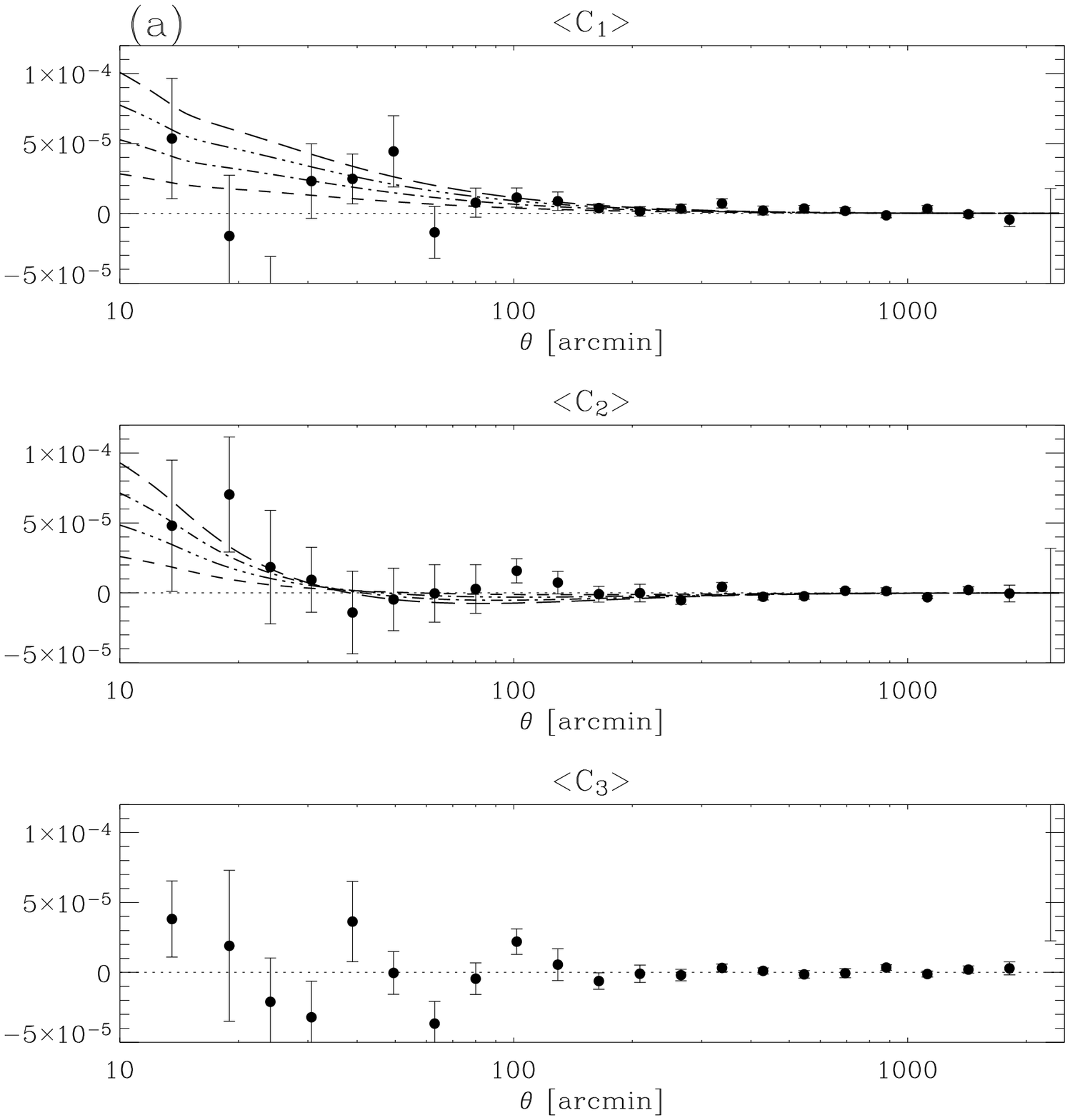}
\includegraphics[scale=0.42]{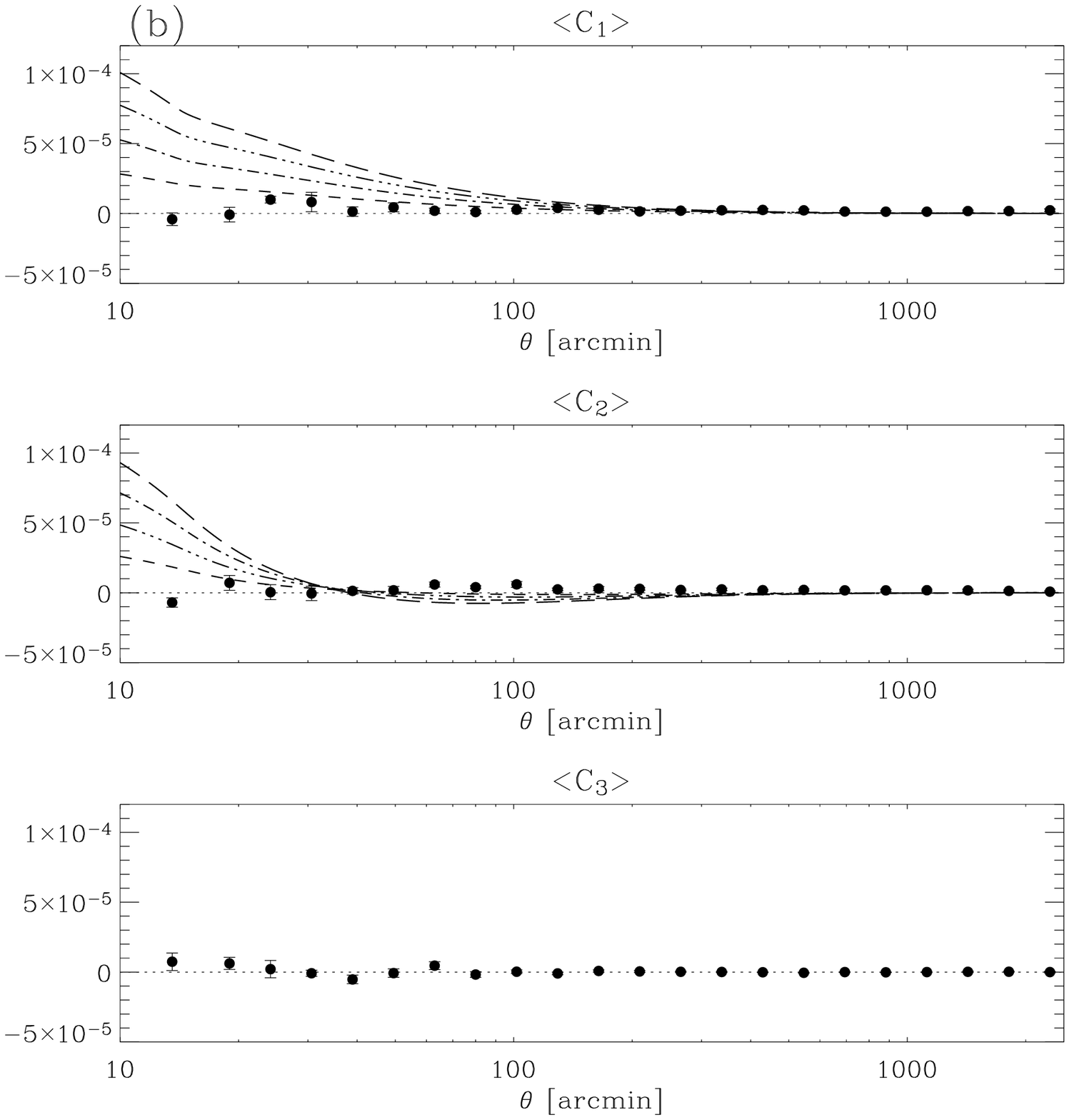}
\caption{(a) The shear correlation function of the data after HA-DEC
  corrections.The lines over-plotted
  indicate predictions from the $\Omega_m =0.3$ $\Lambda$CDM model, with
  $\sigma_8 = 0.9$, and source redshifts lying on the plane of $z_m=
  1, 1.5, 2, 2.5$, respectively. (b) The shear correlation functions of the simulation after
  HA-DEC corrections. The corrections follow the exact same procedure
  as the corrections to the data. The lines show the $\Lambda$CDM
  predictions as in (a).
\label{fig:corre_bins}
}
\end{figure*}

\subsection{Corrected Shear Correlation Functions}
\label{corrections}

First, let us re-examine the averaged shear pattern on the HA-DEC
plane (Fig. \ref{fig:data_hadec}a).  It is evident that this
large-scale pattern is contributed almost entirely by systematics.
Using the averaged data, we fit a two-dimensional 2nd-order polynomial
to this plane, and subsequently subtract the fitted systematics from
individual sources. The sources are grouped according to their sizes
and observed l-m position in the field for the modeling and the
corrections.  We have tried both polynomial fits and spline fits to
the systematics and find that the resulting residuals are comparable.
The residual shear pattern is shown in Fig.
\ref{fig:data_hadec_correct} and appears small in amplitude and random
in orientation, as desired. We discard measurements with averaged
residual shear values greater than 1.5\% in all following analyses.
To test for our control of the systematics, we perform the same
procedure with the simulated shears; the residual shear pattern is
random and very small in amplitude.

We then study the shear correlations function after the HA-DEC
corrections for both the data and the simulation.  As shown in
Fig. \ref{fig:corre_bins}a, the amplitudes of the data $C_1$ and $C_2$
correlations are an order-of-magnitude smaller than those of the
uncorrected data (Fig. \ref{fig:data_correlation}a), and the high
amplitude tail of $C_3$ on large angular scales seen before the correction
has disappeared. The shear correlation functions of the simulation
are also well-behaved on large angular scales, but have much smaller
amplitudes between 10 and $100^{\prime}$ (see Fig. \ref{fig:corre_bins}b).

The correction for the $l$-$m$ dependence is more complicated.  After
source coaddition, which depends on the declination of the phase
center and the number of times a source was observed, for most sources
the $l$-$m$ shear pattern roughly cancels out, while for others it is
enhanced or distorted. This is potentially a contamination to the
lensing signal that is difficult to isolate and correct for.  We
nevertheless perform a correction on the $l$-$m$ plane similar to that
for the HA-DEC plane. First, we examine the shear pattern on the
$l$-$m$ plane using observed and simulated shears that have been
corrected for the HA-DEC pattern, fit a low-order polynomial to the
plane, and then subtract the fitted pattern from individual
measurements. The sources are grouped according to their size for
modeling and corrections, and measurements with averaged residual
shear values greater than 1.5\% are discarded.  After corrections and
coaddition, we again compute the correlation functions for both the
data and the simulation.  Interestingly, the correlation functions are
almost identical to those that have been corrected for HA-DEC patterns
only; i.e., the $l$-$m$ pattern correction does not alter the
correlation functions of the data or of the simulation. We also
checked our corrections by considering subsamples with different
source sizes and did not find significant differences in the resulting
correlation functions.

We also ran a further series of correlation function tests. We consider
using only sources within $15'$ of the phase-tracking center where the
$l$-$m$ distortions are small, excluding sources at low declination
($<10^{\circ}$) where the combined HA-DEC and $l-m$ distortions are
severe, relaxing or strengthening our imposed constraints on excluding
source close companions (discarding sources within $5''$ to $60''$ of
one another), and changing the range in major axes of the selected
sources. Remarkably, the shear correlation functions of the data and of
the simulation remain very similar to those of
Figs. \ref{fig:corre_bins}a and \ref{fig:corre_bins}b, respectively,
in every case.

We therefore conclude that the HA-DEC corrected shear correlation
functions are fairly robust, and that the systematics dependence on
the four parameters, HA, DEC, $l$ and $m$ are well-reproduced by the
simulations.  We thus take the simulated shear correlation functions
after HA-DEC corrections, which are non-zero on most scales, as our
best estimation for the residual systematics in the HA-DEC and $l$-$m$
parameter space.  To correct for these, we subtract the residual
simulated $C_1$ and $C_2$ correlation functions from the corresponding
data correlation function. The errors in each correlation function are
then added in quadrature.

\section{Results}
\label{results}

\subsection{$M_{ap}$ Statistics}

\begin{figure}[t]
\centering
\includegraphics[scale=0.42]{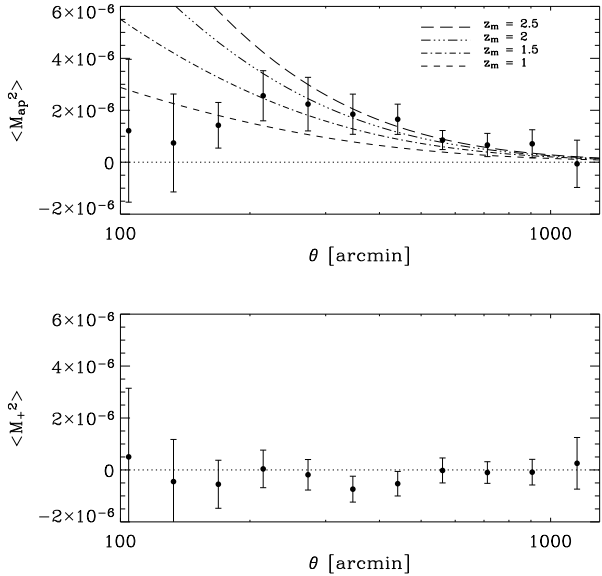}
\figcaption{$M_{ap}$ statistics as a function of aperture radius
$\theta$.  The top panel shows the E-mode variance $\langle M_{ap}^2
\rangle$, while the bottom panel shows the B-mode variance $\langle
M_{\perp}^2 \rangle$. The curves indicate predictions from the
$\Omega_m =0.3$ $\Lambda$CDM model, with $\sigma_8 = 0.9$, $\Gamma=0.21$,
and several source median redshifts $z_m$. \label{fig:map_bins}
}
\end{figure}

Using the shear correlation functions $C_1(\theta)$ and $C_2(\theta)$
measured above, we compute the $M_{ap}$ 2-point statistics using
Eq. (\ref{eqn:map}). The variances $\langle M_{ap}^2 \rangle$ and
$\langle M_{\perp}^2 \rangle$ contain the E-mode and B-mode,
respectively, and are plotted in Fig. \ref{fig:map_bins} as a function
of aperture radius $\theta$. While the E-mode displays a
significant signal, the B-mode amplitude is consistent with 0 on all
scales. The scale dependence of the E-modes for $\theta \gtrsim 200'$
is consistent with that expected for the $\Lambda$CDM model (see
curves), but deviates from this model on smaller scales. The
$M_{ap}$ statistics with an aperture radius of $\theta$ is mostly
sensitive to angular scales of about $\theta / 4$. The depleted
E-modes may therefore reflect the presence of systematics on scales
smaller than $50'$, which roughly corresponds to the scale of the
individual fields ($40'$ in diameter).  This could be due to an over-
or under-correction of one or several of the systematic effects
discussed in section \S\ref{corrections}.

\begin{figure}[t]
\centering
\includegraphics[scale=0.42]{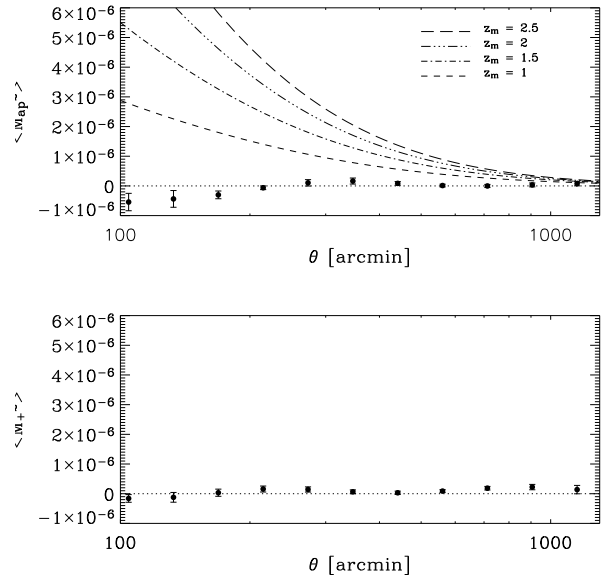} 
\figcaption{Same as the previous figure,
but the $M_{ap}$ statistics are computed using the simulations, which
serve as a null test.
\label{fig:map_sim}
}
\end{figure}

In addition to the checks described in \S\ref{corrections}, we divided
the data into high- and low-latitude halves, and east and west
hemispheres, and computed again the $M_{ap}$ statistics in each case;
the results are consistent with that of Fig. \ref{fig:map_bins}: the
lensing signals are not confined to any particular part of the sky.
Similarly, we separated the sources into two groups according to the
observed flux density; again, the $M_{ap}$ statistics do not differ
significantly. As a comparison, the $M_{ap}$ statistics computed from
the simulations are shown in Fig. \ref{fig:map_sim}.  The E- and
B-modes of the simulation are consistent with zero on most scales, as
expected, with the exception of slightly negative E-mode values on
scales smaller than $\sim 200'$ ($50^{\prime}$ in real space).  This
suggests a possible contamination from systematics on small angular
scales, as discussed above.

\begin{figure}[t]
\centering
\includegraphics[scale=0.42]{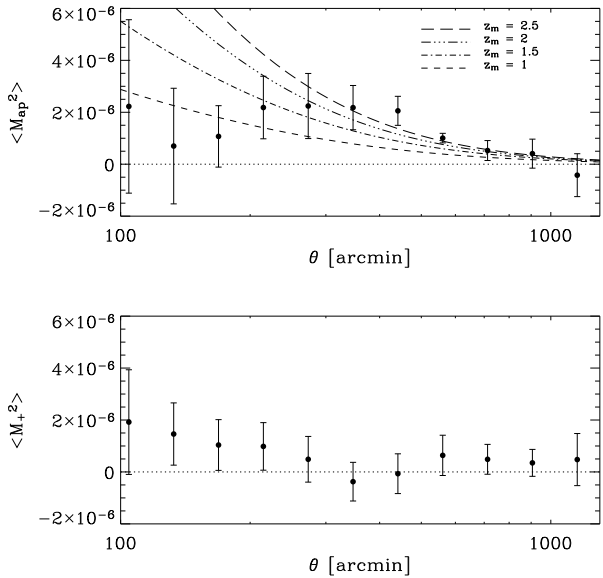}
\figcaption{Same as Fig. \ref{fig:map_bins}, but this time for only those 
sources lacking optical counterparts in the APM catalog.
\label{fig:map_bins_opt}
}
\end{figure}

In order to test for the presence of a lensing signal on scales larger
than $50'$, we removed FIRST sources for which an optical counterpart
could be identified in the APM catalog (McMahon \& Irwin 1992), which
amounts to 10\% of the FIRST sample.  Excluding quasars, which are
both rare and mostly point sources (all of which are excluded from our
source sample), the redshifts of the APM sources peak at around $z
\sim 0.15$ and are for the most part bounded by $z \lesssim 0.3$
(McMahon et al.  2002).  By excluding them, we increase the median
redshift of the source sample and thus expect the lensing signal to
increase. Fig.  \ref{fig:map_bins_opt} shows the $M_{ap}$ statistics
for this new sample.  The E-mode signal has a significance of
$3.6\sigma$, as measured at the $\theta \sim 450'$ scale.  Compared to
Fig.  \ref{fig:map_bins}, the E-mode signal on scales $300' \lesssim
\theta \lesssim 700'$ is indeed larger by 10-20\%.  This confirms the
presence of the lensing signal on these scales.

As an interesting exercise, we calculate the median redshift derived
from the DP redshift models (see \S\ref{zdist}) by excluding the $z <
0.3$ region; the various median redshifts shown in Fig. \ref{fig:dndz}
increase by about 10-15\%.  Since the $M_{ap}$ lensing signal
increases roughly as $z_m^2$ to first order, the changes in the
measured lensing signal wrought by the exclusion of the low-redshift
sources is consistent with that expected from the consequent change in
the estimated $z_m$ from the models.

\subsection{Cosmological Implications}
\label{cosmology}
Using the $M_{ap}$ statistics from the sample without optical
counterparts, we fit cosmological models to our data by computing the
$\chi^2$ values:
\begin{equation}
\chi^2= ({\mathbf d}-{\mathbf m})^{T} {\mathbf W}^{-1} ({\mathbf
d}-{\mathbf m}),
\end{equation}
where ${\mathbf d}$ is the $M_{ap}$ data vector, ${\mathbf m}$ is the
$\Lambda$CDM model vector and ${\mathbf W}$ the covariance matrix.
For ${\mathbf d}$ we use the $M_{ap}$ values with $\theta>200'$, thus
avoiding sub-pointing scales which have unreliable systematics
correction (see discussion above). The covariance matrix is computed
from the field-to-field variations as described in \S\ref{correlation}
and is given by
\begin{equation}
W_{ij}={\Sigma_n ({d_i}_n - \langle d_i \rangle)({d_j}_n - \langle d_j
\rangle) \over N(N-1)},
\end{equation}
where the sum is over different subsamples $n$, and $N=12$ is the total
number of subsamples.  The $d_i$'s are the $M_{ap}$ E-mode values at
different scales, and $\langle d_i \rangle$ is their average over all
12 fields. Since the B-modes quantify the contamination from
systematics and are consistent with zero at the selected angular scales,
we add the B-mode covariance to the E-mode covariance matrix. As a
result, the covariance matrix accounts for all the sources of errors,
namely, intrinsic source shapes, shape measurement errors, cosmic
variance, and systematics.

The data points from angular scales $200'<\theta<1000'$ are partially
degenerate and contain only 3-4 independent measurements.  We
therefore use a singular value decomposition to calculate ${\mathbf
W}^{-1}$, and discard the singular values of ${\mathbf W}$ which are
negligible compared to the rest. Since the $M_{ap}$ data vector was
initially computed at small angular intervals, we only keep four data
points which contain the highest signal-to-noise ratios and are
independent.

The redshift distribution of our sample is rather uncertain, as
indicated by the various models for the radio source redshift-distribution
discussed in \S\ref{zdist}.  We therefore chose to vary two parameters
in our fit: $\sigma_8$, the mass power spectrum normalization in
spheres of $8 h^{-1}$ Mpc, and $z_m$, the median source redshift. We
assumed a $\Lambda$CDM model with fixed values of $\Omega_m = 0.3$ and
$\Gamma =0.21$. Note that, although we are probing scales greater than
$8 h^{-1}$ Mpc, the parameter $\sigma_{8}$ is convenient for comparing 
our results with those from other groups and methods. 

\begin{figure}[t]
\centering
\includegraphics[scale=0.42]{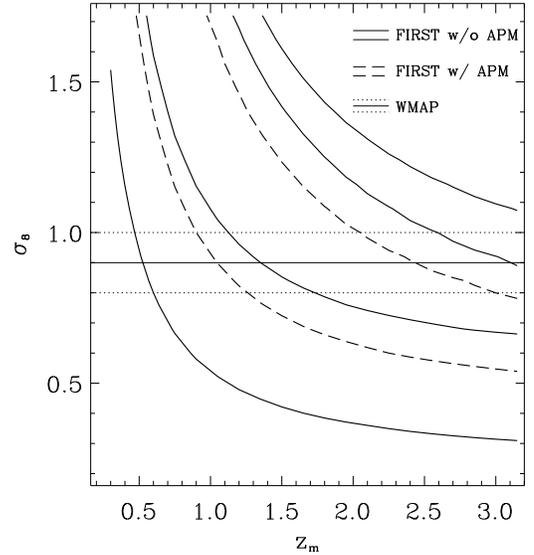} 
\figcaption{ Constraints on the spectrum
normalization, $\sigma_8$, and on the source median redshift, $z_m$,
from our FIRST cosmic shear measurement. The solid lines indicate the
68\% and 95\% CL from FIRST, excluding the predominantly low-redshift
objects with APM optical counterparts. The dashed lines are the 68\%
CL from FIRST including sources with the APM counterparts.  A
$\Lambda$CDM model with $\Omega_m=0.3$ and $\Gamma=0.21$ was
assumed. Also shown for comparison is the constraint from the WMAP CMB
experiment $\sigma_8 = 0.9 \pm 0.1$ (68\%CL; Spergel et al. 2003).
\label{fig:chi2}
}
\end{figure}

The resulting $\chi^2$ contour plot is shown in
Fig. \ref{fig:chi2}. The solid contours indicate the 68.3\%, 95.4\%
confidence levels from FIRST, excluding the predominantly low-redshift
objects with APM optical counterparts.  For comparison, the dashed
contours show the 68.3\% CL constraint from the FIRST sources
including those with the APM counterparts which, as expected, are
consistent with lower $z_m$ values. As a check, we used the linear
power spectrum and the fitting formula from Smith et al. (2003) and
Peacock \& Dodds (1996), and found the contours do not vary
appreciably. This is expected since we are probing the linear part of
the mass power spectrum. To a good approximation, the contours
(excluding APM counterparts) correspond to
\begin{equation}
\sigma_8 \left( \frac{z_m}{2} \right)^{0.6} \simeq 0.95 \pm 0.22,
\end{equation}
where the 68\%CL error includes statistical errors, cosmic variance,
and systematic effects.

Recent cosmic shear measurements in the optical band yield values of
$\sigma_8$ between 0.7 and 1.0 for $\Omega_m \simeq 0.3$ and $\Gamma
\simeq 0.21$ (Bacon et al. 2003; Brown et al. 2003; Hamana et
al. 2003; Hoekstra et al. 2002; Jarvis et al. 2002; Massey et
al. 2003; Refregier, Rhodes, \& Groth 2002; Rhodes, Refregier \& Groth
2004; van Waerbeke et al. 2002). The averaged constraint from several
of these surveys is $\sigma_8 = 0.83 \pm 0.04$ (as compiled by
Refregier 2003) for the same values of $\Omega_m$ and $\Gamma$.  The
constraint, $\sigma_8= 0.9 \pm 0.1$ (68\%CL), from the WMAP CMB
experiment (Spergel et al. 2003) is also shown in Fig. \ref{fig:chi2}.
For source redshifts in the range $1.4 \lesssim z_m \lesssim 3.4$, our
results are consistent (within $1\sigma$) with both of these different
measurements. Reversing the argument and taking the WMAP determination
of $\sigma_8$ as a prior, we find that the median redshift of the
radio sources in our sample (without APM counterpart) is
$z_m=2.2\pm0.9$ at 68\%CL.  This redshift range is also consistent
with the models for the radio source redshift distribution described
in \S\ref{zdist}.

\section{Conclusions} 
\label{conclusion} 

We have presented our cosmic shear measurement using the FIRST Radio
Survey over 8,000 square degrees of sky.  We apply the shear
measurement method described in Chang \& Refregier (2002) to measure
shear directly in Fourier space, where interferometric data are
collected.  In this shapelets approach, we carefully determined the
input parameters for the source shape decomposition, and verified that
the shear estimators are unbiased and robust using realistic
simulations.  The dominant systematic effects associated with
interferometric observations were studied in detail. The analytical
and simulated expectations of the systematics compared well with the
data when considered functions of four observational parameters.  We
corrected the systematics both in this parameter space and in the
correlation function domain, and verified the accuracy of the
corrections with further simulations.

Using the corrected shear correlation functions, we computed the
aperture mass statistic, which decomposes the signal into E- and
B-modes, containing the lensing signal and the non-lensing
contributions, respectively.  We find that the B-modes are consistent
with 0 on all scales considered, while the E-modes displays a
significant lensing signal. The E-mode signal has a significance of
$3.0\sigma$, as measured at the $\theta \sim 450'$ scale. The signal
increases by 10-20\% when sources with optical counterparts (and,
therefore, predominantly at low redshift) are excluded.  This confirms
the presence of a lensing signal which is expected to increase as the
mean source redshift increases.

Using the E-mode $M_{ap}$ statistics on scales $200'< \theta <1000'$
(corresponding to physical scales from about $1^{\circ}$ to
$4^{\circ}$) , we constrained jointly the power spectrum normalization
$\sigma_8$ and the median redshift $z_m$ of our radio source
sample. We find $\sigma_8 (z_m/2)^{0.6} \simeq 1.0\pm0.2$ where the
$1\sigma$ error bars include statistical errors, cosmic variance and
systematics.  A $\Lambda$CDM model with $\Omega_m=0.3$ and
$\Gamma=0.21$ was assumed.  This result is consistent with earlier
determinations of $\sigma_{8}$ from cosmic shear, and with the WMAP
CMB experiment and cluster abundance tests. Taking the prior on
$\sigma_{8}$ from the WMAP experiment, this corresponds to
$z_{m}=2.2\pm0.9$ (68\%CL) for radio sources without optical
counterparts, consistent with existing models for the radio source
luminosity function.

Our shear measurement is complementary to the earlier cosmic shear
results which were all performed in the optical or near-IR band. It is
the first measurement using radio sources observed with
interferometers and is therefore subject to very different systematic
effects. We used a new shear measurement method which allows direct
computation of shear estimators in Fourier space. Moreover, our
measurement probes large angular scales, which are in the linear
regime of structure growth, eliminating the need for uncertain
non-linear corrections in the matter power spectrum. Our measurement
thus offers promising prospects for measuring cosmic shear with future
radio interferometers such as LOFAR and SKA, which offer exciting
opportunities for precision measurements of cosmological parameters
(Schneider 1999).

\acknowledgements{We are grateful to Bob Becker and Rick White for
providing the $uv$ data and the FIRST-APM match catalog, and for
several useful discussions.  We thank Marc Kamionkowski and Arif Babul
for initiating the project and for numerous discussions. We also thank
Siang Peng Oh, Catherine Cress, Zeljko Ivezic, Jacqueline van Gorkom,
and Rick Perley for helpful discussions.  We thank Darren Madgwick for
his help with the tape loading. We also thank the anonymous referee
for numerous constructive comments. TC acknowledges the hospitality of
IoA, Cambridge University and of the Caltech TAPIR group, where part
of the work was done.  At Columbia, this work was supported by NSF
grant AST-98-0273. AR was supported by the EEC TMR network on
Gravitational Lensing and by a Wolfson College Research Fellowship.
This work was performed on the UK-CCC COSMOS facility, which is
supported by HEFCE and PPARC and conducted in cooperation with Silicon
Graphics/Cray Research utilizing the Origin 3800 supercomputer. }

\appendix

\section{Effect of Sensitivity Variations on Source Shapes}
In general, the sensitivity of the image is not exactly constant
across the image. Here, we compute the effect of the sensitivity
variations on the shapes of sources. We then apply our
results to the case of a radial primary beam pattern affecting a
circular Gaussian source.

Let $i({\mathbf x})$ be the intrinsic intensity of a source as a
function of position ${\mathbf x}$ on the image. The observed
intensity is $i'({\mathbf x}) = i({\mathbf x}) s({\mathbf x})$, where
$s({\mathbf x})$ is the sensitivity function. In the application
below, $s({\mathbf x})$ is simply the primary beam function. The
true centroid ${\mathbf x^{0}}$ of the source is defined by
\begin{equation}
{\mathbf x^{0}} = \int d^{2}x ~x_{i} i({\mathbf x}) \left/ 
  \int d^{2}x ~i({\mathbf x}) \right. .
\end{equation}
We will assume that the source angular size is small compared to the
scale on which $s({\mathbf x})$ varies. By 
expanding $s({\mathbf x})$ in a Taylor series about ${\mathbf x^{0}}$,
it is easy to show that the observed centroid position ${\mathbf
x^{0\prime}}$ is, to lowest order,
\begin{equation}
{\mathbf x^{0\prime}} \simeq {\mathbf x^{0}} + J_{ij} S_{j},
\end{equation}
where the summation convention was used and $J_{ij}$ is the normalized
quadrupole moment of the source defined as $J_{ij} \equiv \int d^{2}x ~(x_{i}-x_{i}^{0}) 
  (x_{j}-x_{j}^{0}) i({\mathbf x}) / 
  \int d^{2}x ~i({\mathbf x})$. We
have also defined the normalized derivative tensors of the sensitivity
function as
\begin{equation}
S_{i} \equiv \frac{\partial s({\mathbf x^{0}})}{\partial x_{i}} 
  \left/ s({\mathbf x^{0}}) \right. ,~~~~
S_{ij} \equiv \frac{\partial^{2} s({\mathbf x^{0}})}
 {\partial x_{i} \partial x_{j}} \left/ s({\mathbf x^{0}}) \right.
\end{equation}
We can also compute the observed normalized quadrupole moments of the
source
\begin{equation}
J_{ij}' \equiv \int d^{2}x ~(x_{i}-x_{i}^{0\prime}) 
  (x_{j}-x_{j}^{0\prime}) i'({\mathbf x}) \left/ 
  \int d^{2}x ~i'({\mathbf x}) \right. ,
\end{equation}
about the observed centroid position ${\mathbf x^{0\prime}}$.  To
lowest order, we find
\begin{equation}
J_{ij}' \simeq J_{ij} + S_{k} J_{ijk} - \frac{1}{2} S_{kl} J_{kl} J_{ij}
  + \frac{1}{2} S_{kl} J_{ijkl} - S_{k} S_{l} J_{jk} J_{jl},
\label{eq:jij_sens}
\end{equation}
where $J_{ij}$,$ J_{ijk}$ $J_{ijkl}$ are the true normalized second,
third and fourth order moments of the source about ${\mathbf
x^{0}}$. The terms
in $S_{kl}$ arise from the second order derivative of the sensitivity
function which distorts the image. The terms in $S_{k}$ arise from
the shift from the true centroid.

We will now focus on the case of a radial primary beam pattern for
which the sensitivity $s({\mathbf x}) = S(x)$ is only a function of
the radius $x$ from the center of the image. In this case, $S_{i} =
\frac{s'}{s} \hat{x}_{i}$, and $S_{ij} = \frac{s''}{s} \hat{x}_{i}
\hat{x}_{j} + \frac{s'}{sx} (\delta_{ij} - \hat{x}_{i} \hat{x}_{j})$,
where $\hat{x}_{i} \equiv x_{i}/x$ is the unit radial vector, and $s'$
and $s''$ are derivatives of $s$ with respect to $x$.  For simplicity
we consider a Gaussian source of rms radius $r$, such that $i(x)
\propto e^{-\frac{x^{2}}{2 r^{2}}}$. In this case, the true moments
are simply given by $J_{ij} = \delta_{ij} r^{2}$, $J_{ijk}=0$, and
$J_{1111}=J_{2222}=3 r^{4}$, $J_{1122}=r^{4}$ and $J_{1112}
=J_{1222}=0$. By applying these specific expressions to
equation~(\ref{eq:jij_sens}), we can find the observed ellipticity
$\epsilon_{i}' \equiv \{ J_{11}' - J_{22}', 2 J_{12}' \} / (J_{11}' +
J_{22}')$ and square radius $r^{2\prime} \equiv \frac{1}{2} (J_{11}' +
J_{22}')$ of the source. We find
\begin{equation}
\label{eqn:epbeam}
\epsilon_{i}' = \frac{r^{2}}{2s} \left[s'' -\frac{s'}{x} -
\frac{(s')^{2}}{s} \right] \hat{\epsilon}_{i}^{r}
\end{equation}
and 
\begin{equation}
r^{2\prime} = r^{2} + \frac{r^{4}}{2s} \left[s'' +\frac{s'}{x} -
\frac{(s')^{2}}{s} \right],
\end{equation}
where $\hat{\epsilon}_{i}^{r} = \{ x_{1}^{2} - x_{2}^{2}, 2 x_{1}
x_{2} \}/(x_{1}^{2} + x_{2}^{2})$ is the unit radial ellipticity
vector. Thus, the observed ellipticity pattern will be radial
(tangential) if the quantity $\left[s'' -\frac{s'}{x} -
\frac{(s')^{2}}{s} \right]$ is positive (negative). In section
\S\ref{pbeam}, we apply these results to the specific case of the
VLA primary beam.

\end{document}